\documentclass[a4paper,11pt]{article}
\pdfoutput=1 % if your are submitting a pdflatex (i.e. if you have
\usepackage{jheppub} % for details on the use of the package, please
\usepackage[T1]{fontenc} % if needed

\usepackage{pgfplots}
\usepackage{tikz}
\title{\boldmath $T\overline{T}$ deformations as TsT transformations}

\author[a]{Alessandro Sfondrini,}
\author[b]{Stijn J.\ van Tongeren}

\affiliation[a]{Institut f\"ur theoretische Physik, ETH Z\"urich\\
Wolfgang-Pauli-Strasse 27, 8093 Z\"urich, Switzerland}
\affiliation[b]{Institut f\"ur Physik, Humboldt-Universit\"at zu Berlin,\\
IRIS Geb\"aude, Zum Grossen Windkanal 6, 12489 Berlin, Germany}

\emailAdd{sfondria@itp.phys.ethz.ch}
\emailAdd{svantongeren@physik.hu-berlin.de}

\abstract{The relationship between $T\overline{T}$ deformations and the uniform light-cone gauge, first noted in arXiv:\href{https://arxiv.org/abs/1804.01998}{1804.01998}, provides a powerful generating technique for deformed models.
We recall this construction, distinguishing between changes of the gauge frame, which do not affect the theory, and genuine deformations. We investigate the geometric interpretation of the latter and argue that they affect the global features of the geometry before gauge fixing. Exploiting a formal relation between uniform light-cone gauge and static gauge in a T-dual frame, we interpret such a change as a TsT transformation involving the two light-cone coordinates. In the static-gauge picture, the $T\overline{T}$ CDD factor then has a natural interpretation as a Drinfel'd-Reshetikhin twist of the worldsheet S~matrix.
To illustrate these ideas, we find the geometries yielding a $T\overline{T}$ deformation of the worldsheet S~matrix of pp-wave and Lin-Lunin-Maldacena backgrounds.
}

\newcommand{\de}{\text{d}}

\newcommand{\AdS}[1]{\mathrm{AdS}{}_{#1}}
\newcommand{\CFT}[1]{\mathrm{CFT}{}_{#1}}

\newcommand{\psu}{\mathfrak{psu}}
\newcommand{\su}{\mathfrak{su}}
\renewcommand{\u}{\mathfrak{u}}

\newcommand{\so}{\mathfrak{so}}

\begin{document} 
\maketitle
\flushbottom

\section{Introduction}

The study of two dimensional quantum field theories (QFTs) plays an important role in our understanding of condensed-matter system, of string theory---where the string worldsheet is two-dimensional---and of QFT in general, providing useful toy models that may capture interesting physical features of higher-dimensional theories. Even among two-dimensional models, only some rather special theories can be understood in full detail, usually because they enjoy additional symmetries such as \textit{conformal invariance} or \textit{integrability}. Given such an exactly solvable theory, it is interesting to try and deform it in such a way as to maintain its exact solvability. A rather general class of such deformations can be constructed out of the conserved currents of a theory. Perhaps the most famous example is the \textit{marginal} deformation of a conformal field theory (CFT) by a composite operator constructed out of one chiral and one anti-chiral current---a $J\overline{J}$ deformation. \textit{Relevant} deformations of CFT are also interesting, as they generate a renormalisation group flow and can give rise to families of integrable theories.

More recently, \textit{irrelevant} deformations have been considered, most notably the so-called $T\overline{T}$ deformation. This deformation can be constructed for any two-dimensional Poincar\'e-invariant QFT---conformal, integrable, or not---and it is sourced by the determinant of the stress-energy tensor, 
$\det[T_{\alpha\beta}]=T_{00}T_{11}-T_{01}T_{10}$~\cite{Zamolodchikov:2004ce}.
Interestingly, this deformation acts in a simple way on the spectrum of the original theory: each energy level evolves according to an ordinary differential equation (ODE)~\cite{Smirnov:2016lqw,Cavaglia:2016oda}. In a similar way, the classical Hamiltonian and Lagrangian obey an ODE in the space of fields, which can be often solved in closed form~\cite{Cavaglia:2016oda,Bonelli:2018kik}.  
Over the last three years, $T\overline{T}$ deformations of a number of integrable~\cite{Cavaglia:2016oda, Conti:2019dxg,Conti:2018jho}, as 
well as of more general~\cite{Chen:2018eqk,Aharony:2018bad,Cardy:2018jho,Araujo:2018rho}  theories have been considered.%
\footnote{%
Interesting applications to several classes of two-dimensional theories, such as supersymmetric theories~\cite{Baggio:2018rpv,Chang:2018dge,Jiang:2019hux, Chang:2019kiu,Cribiori:2019xzp}, 2-d gravity~\cite{Dubovsky:2017cnj,Dubovsky:2018bmo,Conti:2018tca,Ishii:2019uwk} and AdS$_3$/CFT$_2$ holography~\cite{McGough:2016lol,Giveon:2017nie,Giveon:2017myj, Asrat:2017tzd, Baggio:2018gct,Dei:2018mfl,Gorbenko:2018oov,Dei:2018jyj, Giveon:2019fgr,Giribet:2017imm}, also emerged.}

A particularly striking link emerged between string theory and $T\overline{T}$ deformations, fueled by the initial observation that the $T\overline{T}$ deformation of a theory of free bosons is related to strings in flat space~\cite{Cavaglia:2016oda}, see also refs.~\cite{Dubovsky:2012wk,Caselle:2013dra}.
It was subsequently understood~\cite{Baggio:2018gct} that the link between strings and $T\overline{T}$ deformations is much more general and becomes particularly transparent in the \textit{uniform light-cone gauge} of refs.~\cite{Arutyunov:2004yx,Arutyunov:2005hd,Arutyunov:2006gs}, see also ref.~\cite{Arutyunov:2009ga} for a pedagogical review.
In fact, this framework can be used as a powerful technique to generate $T\overline{T}$-deformed actions: finding the deformed Hamiltonian requires solving an \textit{algebraic} equation rather than an ODE~\cite{Baggio:2018rpv,Frolov:2019nrr}. Moreover, this approach can be even applied to much more general current-current deformations~\cite{Frolov:2019xzi}, of the type considered in refs.~\cite{Guica:2017lia,Bzowski:2018pcy,Nakayama:2018ujt,Chakraborty:2018vja, LeFloch:2019rut, Guica:2019vnb,Chakraborty:2019mdf}.

It is this link with string theory in uniform light-cone gauge that is the focus of this paper. In a nutshell, given a two-dimensional model with~$D$ fields which we want to deform, we uplift it to a reparametrisation-invariant model with $D+2$ fields by adding two ``longitudinal'' coordinates---two bosonic fields $X^\pm$ invariant under shifts~$X^\pm\to X^\pm+\delta X^\pm$. Then, gauge fixing yields back the original model. In uniform light-cone gauge, a change of \textit{gauge frame} mimics the $T\overline{T}$ of the original Hamiltonian density. This can be seen from the fact that the volume~$R$ of the gauge-fixed model depends on the gauge-frame parameter~$a$~as
\begin{equation}
    R= R_0 +a\,H_{\text{w.s.}}\,,
\end{equation}
where $H_{\text{w.s.}}$ is the energy of the two-dimensional theory (in string-theory language, the \textit{worldsheet Hamiltonian}). This is the same dependence observed in $T\overline{T}$-deformed theories~\cite{Smirnov:2016lqw,Cavaglia:2016oda}. 
The Hamiltonian density of $H_{\text{w.s.}}$ also depends on~$a$, precisely in such a way that the $a$ dependence cancels in physical quantities such as the spectrum.

It is then important to distinguish between gauge-frame changes, that \textit{do not} affect a theory, and genuine deformations. In the latter the change of the  Hamiltonian density \textit{is not compensated} by a redefinition of the worldsheet length, and hence the spectrum changes like in~$T\overline{T}$ deformation. We consider this case and investigate the effect of such a deformation \textit{on the uplifted geometry}. We will argue that such a deformation does not affect the geometry \textit{locally}, but does so \textit{globally}. We will see that, exploiting a formal relation between uniform light-cone gauge and static gauge~\cite{Kruczenski:2004cn}, we can make the geometric interpretation of the deformation more transparent, and recast it as a T-duality--shift--T-duality (TsT) transformation~\cite{Lunin:2005jy} involving the two longitudinal coordinates. Indeed, in a string sigma model, such TsT transformations can equivalently be understood as a twist of the boundary conditions of the involved coordinates~\cite{Frolov:2005dj,Alday:2005ww,vanTongeren:2018vpb}, rather than a genuine modification of the local geometry. For integrable models, such a twist of the boundary conditions results in a twist of the Bethe-Yang equations~\cite{Beisert:2005if}. Equivalently, from the point of view of the deformed geometry, a TsT transformation leads Drinfel'd-Reshtikin twist~\cite{Drinfeld:1989st,Reshetikhin:1990ep} of the worldsheet S~matrix~\cite{Ahn:2010ws,Ahn:2012hs}. Taking this view, we can interpret the CDD factor~\cite{Castillejo:1955ed} arising from $T\overline{T}$ deformation~\cite{Smirnov:2016lqw,Cavaglia:2016oda} as such a Drinfel'd-Reshtikin twist on the Cartan charges corresponding to the two longitudinal directions.
This reinforces the identification between $T\overline{T}$ deformations and gauge fixing. In fact, the $T\overline{T}$ CDD factor can be taken as a \textit{definition} of such a deformation~\cite{Cavaglia:2016oda}.

We can apply these ideas to explicitly construct integrable deformations of superstrings backgrounds of particular interest. The resulting geometry will be such that, once a light-cone gauge is fixed, its worldsheet S~matrix differs from the underformed one precisely by the $T\overline{T}$ CDD factor. In the case of AdS$_5\times$S$^5$, we can construct a string background which yields a  $T\overline{T}$ deformation of Beisert's S~matrix~\cite{Beisert:2005tm} in the ``string frame'' of refs.~\cite{Arutyunov:2006yd}; this will preserve integrability by virtue of being a $T\overline{T}$ deformation.%
\footnote{%
It would be interesting to understand what the gauge-theory construction dual to such a background might be. We will briefly speculate on this point in section~\ref{sec:speculations}.
}
It is also possible to consider such a deformation in the case where the original geometry is not integrable though of course the resulting spectral problem will be much less tractable. As an illustration we consider Lin-Lunin-Maldacena (LLM) backgrounds,%
\footnote{%
The (non-)integrability of LLM geometries is discussed in ref.~\cite{Chervonyi:2013eja}.
}
where the deformation has a particularly clean interpretation.

This paper is structured as it follows. In section~\ref{sec:lcgauge} we review is some detail the uniform light-cone gauge and its relation with $T\overline{T}$ deformations. In section~\ref{sec:deformed-bgd} we discuss the geometrical interpretation of such deformations, the relation to TsT transformations, and the interpretation of the CDD factor as a Drinfel'd-Reshtikin twist. In sections~\ref{sec:example} and \ref{sec:LLM} we exemplify our arguments for pp-wave and LLM backgrounds, respectively. We also briefly speculate on the gauge-theory interpretation of a deformation of AdS$_5\times$S$^5$, relegating some bulky expression to appendix~\ref{sec:appexplicit}. We present some concluding remarks in section~\ref{sec:conclusion}. 
Our results can be straightforwardly generalised to the case of current-current deformations involving a $\mathfrak{u}(1)$ current~$J$, such as $J\overline{T}$ or~$T\overline{J}$ deformations; we briefly discuss how in appendix~\ref{sec:TJappendix}.

\section{\texorpdfstring{$T\overline{T}$}{TTbar} deformations and uniform light-cone gauge}
\label{sec:lcgauge}

The relationship between $T\overline{T}$  deformations and uniform light-cone gauge%
\footnote{The uniform light-cone gauge was introduced in refs.~\cite{Arutyunov:2004yx,Arutyunov:2005hd,Arutyunov:2006gs} and has been reviewed in detail in ref.~\cite{Arutyunov:2009ga}.}
has been first noted in ref.~\cite{Baggio:2018gct} and subsequently exploited to construct $T\overline{T}$-deformed Lagrangians, see ref.~\cite{Baggio:2018rpv} and in particular~\cite{Frolov:2019nrr,Frolov:2019xzi}. Even if this construction is fairly well-known in the literature, let us briefly review it for the sake of being self-contained.

\subsection{Uniform light-cone gauge}

We consider a non-linear sigma model with metric $G_{\mu\nu}(X)$, where $X$ collectively denotes all the fields, and B-field $B_{\mu\nu}(X)$. The metric part of the action is coupled to a two-dimensional metric $\gamma^{\alpha\beta}$, which we take to have unit determinant. By construction, the theory is invariant under re-parametrisations at the classical level. For the moment we will be interested in the classical theory, and we will not assume that the metric and B-field describe a string background. We will however assume that the metric has at least two shift isometries: one for a time-like coordinate which we denote by $t$, $t\to t+\delta t$, and which yields the target-space energy $E$, and one for $\phi\to\phi+\delta \phi$, which yields some (angular) momentum $J$.

All in all we have
\begin{equation}
S=-\frac{1}{2}\int\limits_{-\infty}^{+\infty}\de\tau\int\limits_{0}^{R}\de\sigma \Big(
\gamma^{\alpha\beta}\partial_\alpha X^\mu\,\partial_\beta X^\nu\,G_{\mu\nu}(X)+\varepsilon^{\alpha\beta}\partial_\alpha X^\mu\,\partial_\beta X^\nu\,B_{\mu\nu}(X)
\Big).
\end{equation}
The minus sign takes into account that the worldsheet metric has signature $(-,+)$.
It is convenient to introduce the momenta $p_\mu$, which are canonically conjugated to $X^\mu$:
\begin{equation}
p_\mu = \frac{\delta S}{\delta \partial_\tau X^\mu}=-\gamma^{0\beta}\partial_{\beta}X^\nu G_{\mu\nu}(X)-\acute{X}^\nu B_{\mu\nu}(X)\,,
\end{equation}
where we introduced the notation $\acute{X}^\nu\equiv \partial_\sigma X^\mu$. By Noether's theorem we immediately get two conserved charges
\begin{equation}
E= -\int\limits_{0}^{R}\de\sigma\ p_t\,,\quad \text{and} \quad
J= \int\limits_{0}^{R}\de\sigma\ p_\phi\,.
\end{equation}
An advantage of the first order formalism is that the action takes the form
\begin{equation}
\label{eq:firstorderaction}
S=\int\limits_{-\infty}^{+\infty}\de\tau\int\limits_{0}^{R}\de\sigma 
\Big(
p_\mu \dot{X}^\mu +\frac{\gamma^{01}}{\gamma^{00}}\mathcal{C}_1+\frac{1}{2\gamma^{00}}\mathcal{C}_2
\Big),
\end{equation}
where the worldsheet metric takes the form of a Lagrange multiplier and yields the two Virasoro constraints:
\begin{equation}
\label{eq:Virasoro}
\begin{aligned}
&0=\mathcal{C}_1=p_\mu\acute{X}^\mu\,,\\
&0=\mathcal{C}_2=p_\mu p_\nu\,G^{\mu\nu}+\acute{X}^\mu\acute{X}^\nu\, G_{\mu\nu}
+2G^{\mu\nu}B_{\nu\rho}p_\mu\acute{X}^\rho+ G^{\mu\nu}B_{\mu\rho}B_{\nu\lambda}\acute{X}^{\rho}\acute{X}^{\lambda}
\,,
\end{aligned}
\end{equation}
where we suppressed the dependence of the (inverse) metric and B-field on $X^\mu$.

\paragraph{Choice of the light-cone coordinates.}
We can use the two isometric coordinates $(t,\phi)$ to construct the coordinates that we will use in the gauge fixing. A very natural choice is to define two \textit{light-like} coordinates $X^{\pm}$, by
\begin{equation}
\label{eq:simplelc}
X^{\pm} = \frac{1}{2}\big(\phi \pm t\big)\,.
\end{equation}
However, it is convenient to generalise this choice by introducing two parameters
\begin{equation}
\label{eq:basischange}
X^{+} = a\,\phi+(1-a)\,t\,,\qquad
X^{-} = (1-b)\,\phi-b\,t\,,\qquad \Delta_{ab}\equiv 1-a-b+2ab\neq0\,,
\end{equation}
so that we have
\begin{equation}
p_+= \frac{b}{\Delta_{ab}}p_\phi+\frac{1-b}{\Delta_{ab}}p_t\,,\qquad
p_-= \frac{1-a}{\Delta_{ab}}p_\phi-\frac{a}{\Delta_{ab}}p_t\,.
\end{equation}
Let us note that if $b=1$, then $p_+\sim p_\phi$, independently from $p_t$. We will see below that this case is pathological, so that we shall always assume
\begin{equation}
    \label{eq:bisnot0}
    b\neq1\,.
\end{equation}

\paragraph{Uniform light-cone gauge fixing.}
The uniform light-cone gauge is fixed by the conditions
\begin{equation}
\label{eq:lightconegauge}
    X^+=\tau\,,\qquad p_- = \frac{1}{1-b}\,.
\end{equation}
which identify the worldsheet time $\tau$ with a particular target space direction, $X^+$, and at the same time imposes the momentum density for $p_-$ to be constant. The choice of this constant is a matter of future convenience; for the moment we note that it is compatible with our requirement \eqref{eq:bisnot0}. It is then possible to eliminate the two remaining longitudinal degrees of freedom $X^-$ and $p_+$ by using the Virasoro constraints~\eqref{eq:Virasoro}, obtaining
\begin{equation}
    0=\mathcal{C}_1= p_+\acute{X}^+ + p_-\acute{X}^- + p_i \acute{X}^i
    \quad \Rightarrow \quad \acute{X}^- = - (1-b)p_i\,\acute{X}^i\,.
\end{equation}
while $\mathcal{C}_2=0$ gives a quadratic equation for $p_+$ (which may degenerate into a linear equation should $G^{++}=0$ for some particular choice of $a,b$). We have not found an expression for $X^-$ itself, but only for its $\sigma$-derivative; this is expected, as the action does not depend directly on $X^-$ as this is an isometric coordinate. We have however to require that $X^-$ satisfies appropriate boundary conditions, which we take to be periodic,%
\footnote{It is possible to consider more general boundary conditions, for instance involving winding along $\phi$ is that is a compact coordinate. We refer the reader to refs.~\cite{Arutyunov:2009ga,Frolov:2019nrr} for a discussion of this case.}
\begin{equation}
\label{eq:lvlmatch}
    0=\int\limits_{0}^{R}\de\sigma \acute{X}^- = \int\limits_{0}^{R}\de\sigma \big(- p_i\,\acute{X}^i\big)=P_{\text{w.s.}}\,.
\end{equation}
In the last step we have identified the integral with the total momentum on the worldsheet~$P_{\text{w.s.}}$; Noether's theorem shows that $- p_i\,\acute{X}^i$ is precisely the charge density for the symmetry $\sigma\to\sigma+\delta\sigma$. Indeed~\eqref{eq:lvlmatch} is nothing else but the level-matching constraint.
Finally the action~\eqref{eq:firstorderaction} becomes
\begin{equation}
\label{eq:pplus}
S=\int\limits_{-\infty}^{+\infty}\de\tau\int\limits_{0}^{R}\de\sigma\ 
p_\mu \dot{X}^\mu = 
\int\limits_{-\infty}^{+\infty}\de\tau\int\limits_{0}^{R}\de\sigma\Big(
p_i \dot{X}^i  - (-p_+)\Big)\,,
\end{equation}
where we dropped a total $\tau$-derivative $\dot{X}^-$. Hence we identify the worldsheet Hamiltonian~$H_{\text{w.s.}}$ with $-p_+$,
\begin{equation}
    H_{\text{w.s.}}=-\int\limits_{0}^{R}\de\sigma\ p_+(X^i,\acute{X}^i,p_i)\,,
\end{equation}
which is expected because $H_{\text{w.s.}}$ is canonically conjugated to $\tau$ and hence to $X^+$. As for $p_-$ we find that in this gauge
\begin{equation}
    P_- = \int\limits_{0}^{R} \de\sigma\ p_- =\frac{R}{1-b}\,.
\end{equation}
To conclude, it is useful to make explicit the relations between $H_{\text{w.s.}}, R$ with $E$, $J$ in terms of the parameters $a,b$ introduced in~\eqref{eq:basischange}.
\begin{equation}
\label{eq:HwsR}
    H_{\text{w.s.}}=\frac{(1-b)\, E - b\,J}{\Delta_{ab}}\,,
\qquad
    R= \frac{1-b}{\Delta_{ab}}((1-a)\, J + a\,E)=J + a\, H_{\text{w.s.}}\,.
\end{equation}
We see here that $b=1$ is a singular choice, as we would be matching the worldsheet Hamiltonian with the potentially quantised (angular) momentum~$J$. 
Finally note that, unless $a=0$, the volume~$R$ in which the theory will be quantised will be \textit{state-dependent}, namely it will depend on the energy of each given state. This is of course a first indication of a relation with $T\overline{T}$ deformations, as discussed already in ref.~\cite{Baggio:2018gct} and as we shall review in the next section.

\paragraph{Some choices of the parameters $a,b$.}
Let us briefly comment on some features of this gauge choice which may appear a little unconventional. First of all, the parameter $b$ allows us to change the relation between the worldsheet Hamiltonian~$H_{\text{w.s.}}$ and the target-space energy~$E$. Properly speaking, uniform light-cone gauge corresponds to the choice where $H_{\text{w.s.}}$ is the light-cone Hamiltonian, $H_{\text{w.s.}}= E-J$.%
\footnote{For many string backgrounds, such a choice may preserve some manifest supersymmetry, so that the corresponding vacuum is protected from quantum corrections; this makes the quantisation of the theory substantially simpler.}
This can be simply achieved by setting $b=1/2$:
\begin{equation}
 b=\frac{1}{2}:\qquad   H_{\text{w.s.}}=E-J\,,
\qquad
    R= J+a\,H_{\text{w.s.}}.
\end{equation}
Let us also mention another particular choice of $b$, that is $b=0$, which allows to identify the worlsheet Hamiltonian with~$E$. We then have
\begin{equation}
    b=0:\qquad
    H_{\text{w.s.}}=\frac{E}{1-a}\,,\qquad
    R=J+a\,H_{\text{w.s.}}\,.
\end{equation}
In either case, the choice $a=0$ is particularly simple in that the volume of the theory is fixed in terms of the charge $J$, and hence does not depend on the state---or more precisely, different choices of $J$ yield different superselection sectors that may be studied separately.

\subsection{Changing the gauge frame}
Let us now review the relation between the light-cone gauge introduced above---in particular, the parameter $a$ introduced in~\eqref{eq:basischange}---and $T\overline{T}$ deformations. This was first discussed in \cite{Baggio:2018gct} and subsequently in greater detail in~\cite{Frolov:2019nrr}, building on the extensive existing literature on the uniform light-cone gauge~\cite{Arutyunov:2004yx,Arutyunov:2005hd,Arutyunov:2006gs,Arutyunov:2009ga}.

\paragraph{Changes of gauge frame and the Hamiltonian.}
Varying the parameters $a$ and $b$ introduced in~\eqref{eq:basischange} \textit{must not} have any physical consequence. It is simple to understand this for a variation of $b$, with $a$ fixed. Such a change of course modifies the spectrum of $H_{\text{w.s.}}$. However, it will not affect the spectrum of $E$, defined through~\eqref{eq:HwsR}---it is quite simply a linear redefinition of the operator whose spectrum we are computing. Things are a little more subtle when varying $a$ (keeping $b$ fixed for simplicity). In that case, we have seen that $R$ varies; it is also not hard to see that the Hamiltonian density $-p_+(X^i,\acute{X}^i,p_i)$ also depends explicitly on~$a$. Hence we must have, formally
\begin{equation}
\label{eq:agauge}
    0=\frac{\de}{\de a}H_{\text{w.s.}}= -\frac{\de}{\de a} \int\limits_0^{J+a\,H_{\text{w.s.}}} \de\sigma\ p_+(X^i,\acute{X}^i,p_i;a)\,.
\end{equation}
This property is well-known in the context light-cone gauge-fixed strings~\cite{Arutyunov:2009ga}, and has also been verified perturbatively for a number of models, see \textit{e.g.}~\cite{Klose:2006zd,Sundin:2013ypa,Dei:2018jyj}.

\paragraph{Changes of gauge frame and the S~matrix.}
It is instructive to consider the condition~\eqref{eq:agauge} for theories that whose spectrum can be described in terms of a factorised S-matrix, which is the context where the uniform gauge was originally proposed~\cite{Arutyunov:2006gs}.
This means introducing particles corresponding to the fields~$X^i$ having worldsheet momentum $p$ and energy~$\omega_i(p)$.%
\footnote{The worldsheet momentum $p$ should not be confused with the conjugate momenta~$p_\mu$. The index $i$ denotes the flavour of the particle.} 
Then, the interactions of $H_{\text{w.s.}}$ are captured by the S~matrix, and if this is factorisable it is sufficient to consider the 2-to-2 scattering matrix $S_{i_1i_2}^{i_2'i_1'}(p_1,p_2;a)$, which depends on~$a$, like $H_{\text{w.s.}}$.
The idea is that the energy of a state with momenta $p_1,\dots p_M$ can be computed in the asymptotic states, when all particles are approximately free and
\begin{equation}
    P_{\text{w.s.}} =\sum_{k=1}^M p_k\,,\qquad H_{\text{w.s.}} = \sum_{k=1}^M \omega_{i_k}(p_k)\, .
\end{equation}
In finite volume $R$ the momenta are quantised, as prescribed by the Bethe-Yang equations, which we write down under the assumption that the scattering matrix is diagonal%
\footnote{Non-diagonal S~matrices can diagonalised by nested Bethe ansatz, leading to the same conclusion. Note also that the exact spectrum will also have to account for finite size effects of the type of refs.~\cite{Luscher:1985dn,Luscher:1986pf}, which can be accounted for by the thermodynamic Bethe ansatz~\cite{Yang:1968rm,Zamolodchikov:1989cf}.}
\begin{equation}
    e^{i p_j R(a)} \prod_{k\neq j}^{M} S_{i_j i_k}^{i_k i_j}(p_j,p_k;a) =1\,.
\end{equation}
Already in ref.~\cite{Arutyunov:2006yd} it was argued that the $a$-dependence of the S-matrix should take the form
\begin{equation}
\label{eq:smat}
    S_{i_j i_k}^{i_k i_j}(p_j,p_k;a) =e^{i a \Phi(p_j,p_k)}\,S_{i_j i_k}^{i_k i_j}(p_j,p_k)\,.
\end{equation}
with
\begin{equation}
\label{eq:cdd}
    \Phi(p_j,p_k)= p_k\,\omega_{i_j}(p_j)-p_j\,\omega_{i_k}(p_k)\,.
\end{equation}
This is indeed a ``CDD factor''~\cite{Castillejo:1955ed}, meaning that it solves the homogeneous crossing equation, regardless of the specific form of~$\omega_i(p)$.
Then, using that $P_{\text{w.s.}}=0$ we have
\begin{equation}
    e^{i p_k (J+a H_{\text{w.s.}})} e^{-i a p_k H_{\text{w.s.}}}\prod_{k\neq j}^{M} S_{i_j i_k}^{i_k i_j}(p_j,p_k) =1\,,
\end{equation}
which indeed is $a$-independent.%
\footnote{Strictly speaking the Bethe-Yang equations describe the spectrum only approximately---up to finite-volume corrections. This argument can be straightforwardly repeated at the level of the mirror thermodynamic Bethe ansatz which describes the true finite-volume spectrum.}

\subsection{\texorpdfstring{$T\overline{T}$}{TTbar} deformations vs.\ gauge-frame choices}
Having reviewed some well-established properties of uniform light-cone gauge it is now easy to see the relation with $T\overline{T}$ deformations~\cite{Baggio:2018gct,Frolov:2019nrr}. First of all, the dependence of the volume $R$ on the energy~$H_{\text{w.s.}}$ is precisely such as to reproduce the Burgers equation~\cite{Smirnov:2016lqw,Cavaglia:2016oda}. Secondly, the phase factor~$\Phi(p_j,p_j)$ is precisely the $T\overline{T}$ ``CDD factor'' of refs.~\cite{Dubovsky:2012wk,Caselle:2013dra,Cavaglia:2016oda}. Indeed for a relativistic theory with $p=m\sinh\theta$ and $\omega(p)=m\cosh\theta$ we have $\Phi(p_j,p_k)=m_jm_k\sinh(\theta_k-\theta_j)$.
What is important to note is that the change of gauge frames described above \textit{do not} generate a new theory; indeed we have stressed that a change of $a$ does not affect the spectrum of~$H_{\text{w.s.}}$, see eq.~\eqref{eq:agauge}. What would generate a deformation of the $T\overline{T}$ type is \textit{to deform the Hamiltonian density} $-p_+(X^i,\acute{X}^i,p_i;a)$ by tuning $a$, \textit{without redefining the volume} $R$ accordingly. It is in this sense that the light-cone gauge $a$-dependent frame may be used to generate $T\overline{T}$ deformed Hamiltonian and Lagrangian densities~\cite{Baggio:2018rpv,Frolov:2019nrr}, as well as to study more general deformations~\cite{Frolov:2019xzi}.
In a similar way, a variation of the frame-parameter $b$ also induces a deformation if we vary the Hamiltonian density $-p_+(X^i,\acute{X}^i,p_i;b)$ without changing the relation between $H_{\text{w.s.}}$, $E$ and $J$ of eq.~\eqref{eq:HwsR}.

Our next goal will be to understand such deformations, and in particular those related to $a$, in geometric terms. Let us introduce an \textit{ad-hoc} notation to denote deformations (as opposed to changes of the gauge frame),
\begin{equation}
    a\to \bar{a}=a-\delta a\,,\qquad
    b\to \bar{b}=b-\delta b\,,
\end{equation}
meaning that $\delta a$ and $\delta b$ are \textit{deformation parameters}, which generate genuinely new theories. In particular, the parameter $\delta a$ is proportional to the $T\overline{T}$ deformation parameter.

\section{Deformed backgrounds from \texorpdfstring{$T\overline{T}$}{TTbar}}
\label{sec:deformed-bgd}
In the previous section we reviewed how we can describe the $T\overline{T}$ deformation of a bosonic theory by coupling it to two additional isometric coordinates $t$ and $\phi$ and endowing it with parametrisation invariance. Then the $T\overline{T}$-deformed Hamiltonian (or Lagrangian) density may be obtained from gauge fixing this parent theory and varying the gauge-frame parameter~$a$ while keeping the worldsheet size~$R$ fixed.%
\footnote{More general actions and deformations may be studied in the same way, and we refer the reader to refs.~\cite{Frolov:2019nrr,Frolov:2019xzi} for a detailed discussion of these points.}
A natural question is what is the geometrical interpretation of the deformed parent theory. For instance, let us take a string background, fix uniform light-cone gauge, and then vary the parameters $a,b$ in $-p_+(X^i,\acute{X}^i,p_i;a,b)$ but not in eq.~\eqref{eq:HwsR}. What geometry would lead to such a gauge fixed theory? In order to address this question it will turn out to be convenient to exploit a formal relationship between uniform light-cone gauge and static gauge.

\subsection{\texorpdfstring{$T\overline{T}$}{TTbar} deformations as a coordinate shift}

Let us begin by considering the $T\overline{T}$ deformation in terms of reparametrising the light-cone coordinates. The effect of changing $a$ and $b$ in our light-cone parametrisation amounts to
\begin{equation}
\label{eq:Xpmredef}
    X^+ \rightarrow X^+ + \delta a \, \frac{X^- + (2\bar{b}-1)X^+}{\Delta_{\bar{a}\bar{b}}}, \quad X^- \rightarrow X^- - \delta b \, \frac{X^+ - (2\bar{a}-1)X^-}{\Delta_{\bar{a}\bar{b}}},
\end{equation}
where the $X^\pm$ on the right hand side are our new light-cone coordinates.
It may seem that such a redefiniton is trivial. Indeed this linear map is certainly a local diffeomorphism. Hence \textit{locally} the new metric that we obtain from such a shift will be equivalent to the original one. This does not mean that the geometry will be the same \textit{globally}, unless we also modify the boundary conditions of the field $X^\pm$ according to the shift~\eqref{eq:Xpmredef}, and unless we redefine the interpretation of the charges $P_{\pm}$. Just shifting the coordinates would hence result in a different spectrum for the gauge-fixed theory. It is instructive to work this out in some detail for some examples, such as pp-wave and flat space or AdS$_5\times$S$^5$ and LLM geometries. We will do so in sections~\ref{sec:example} and~\ref{sec:LLM}. Here below we discuss a more general and transparent way to understand the geometric effect of the shift~\eqref{eq:Xpmredef}. To this end, we will exploit a formal relation between the uniform light-cone gauge and the static gauge~\cite{Kruczenski:2004cn}.

\subsection{From uniform light-cone gauge to static gauge}

In the Hamiltonian or first order formalism one fixes a light-cone gauge by fixing $X^+ = \tau$ and $p_- = (1-b)^{-1}$, as in eq. \eqref{eq:lightconegauge}. Alternatively, as shown in \cite{Kruczenski:2004cn}, we can obtain the same result, by T~dualising the action in $X^-$, integrating out the world-sheet metric, and fixing $X^+=\tau$ and the T-dual coordinate $\tilde{X}^-=\sigma/(1-b)$, i.e. fixing a static gauge. Let us briefly review why this is the case. 

To perform T~duality in the $X^-$ direction we gauge the shift symmetry for $X^-$, replacing
\begin{equation}
    \partial_\alpha X^- \rightarrow \partial_\alpha X^- + A_\alpha
\end{equation}
in the Lagrangian, and adding the term $\widetilde{X}^- \epsilon^{\alpha \beta}\partial_\alpha A_\beta$,
\begin{equation}
     L(\partial_\alpha X^+,\partial_\alpha X^-,X^i)
     \to 
     L(\partial_\alpha X^+,\partial_\alpha X^-+A_\alpha,X^i)+\widetilde{X}^- \epsilon^{\alpha \beta}\partial_\alpha A_\beta,
\end{equation}
where the Lagrange multiplier field $\widetilde{X}^-$ ensures that $A_\alpha$ is flat and hence pure gauge. Integrating out $\widetilde{X}^-$ gives back the original Lagrangian, while integrating out $A_\alpha$ gives the Lagrangian of the T-dual model. Upon integrating out $A_\alpha$ we in particular need to take into account the equation of motion for $A_\tau$
\begin{equation}
\label{eq:XmPm}
    \partial_\sigma \widetilde{X}^- = \frac{\partial \mathcal{L}}{\partial \dot{X}^-} = p_-,
\end{equation}
where $p_-$ is the momentum conjugate to the \emph{original} light cone coordinate $X^-$. We see that the gauge condition $p_-=1/(1-b)$ translates to
\begin{equation}
    \widetilde{X}^- = \frac{\sigma}{1-b}\,,
\end{equation} 
in the T-dual picture. The range of $\sigma$ in the T-dual picture is fixed by the requirement that $\widetilde{X}^-$ winds an integer number of times.
This matches with the intuition that T~duality interchanges winding and momentum modes, so that a vacuum with non-zero momentum $P_-$ along $X^-$ has non-zero winding along~$\widetilde{X}^-$.
On the other hand, since we considered no winding along $X^-$ in the original theory, we will have no momentum along~$\widetilde{P}_-$.
To understand the physical meaning of~$\widetilde{P}_-$ we recall that $\tilde{p}_-$ is canonically conjugated to $\tilde{X}^- \sim \sigma$. Indeed using the Virasoro constraint~$\mathcal{C}_1$ we have
\begin{equation}
\label{eq:ptilde}
   0=\mathcal{C}_1=2 \tilde{p}_- + p_i \acute{X}^{i}, \quad \Rightarrow \quad \widetilde{P}_- = \frac{1}{2}P_{w.s.},
\end{equation}
so that a state with zero-winding in the original theory is level-matched in the T-dual description.
In summary, fixing a uniform light-cone gauge is equivalent to T~dualising in $X^-$ and fixing a static gauge instead. This procedure has been applied in setups of increasing generality in \cite{Klose:2006zd,Zarembo:2009au,Arutyunov:2014jfa}.

\subsection{\texorpdfstring{$T\overline{T}$}{TTbar} in the T-dual picture}
Now let us compare light-cone gauge fixing with two different choices of `gauge'  parameter from the T-dual perspective, having in mind to keep $R$ fixed. Starting with a parent theory $\mathcal{T}(a,b)$ with gauge parameters $a$ and $b$, we can T~dualize in $X^-$ to obtain a dual model, $\widetilde{\mathcal{T}}(a,b)$, whose static gauge version is equivalent to the light-cone gauge version of the original. In the parent theory we can vary our choice of gauge parameters, where $a\rightarrow \bar{a} = a-\delta a$ and $b\rightarrow \bar{b} =  b-\delta b$, corresponds to the coordinate redefinition~\eqref{eq:Xpmredef}. In this resulting theory, we can fix a light-cone gauge with respect to our new light-cone gauge coordinates, and again view this from a T-dual perspective. All in all this gives us two theories that in the static gauge are related by a change of the gauge parameters $a$ and $b$:
\begin{equation}
\begin{tikzpicture}[baseline=(current  bounding  box.center)]
\node (T1) at (-4cm,1cm) {${\mathcal{T}}(a,b)$};
\node (T1d) at (-4cm,-1cm) {$\widetilde{\mathcal{T}}(a,b)$};
\node (T2) at (4cm,1cm) {${\mathcal{T}}(\bar{a},\bar{b})$};
\node (T2d) at (4cm,-1cm) {$\widetilde{\mathcal{T}}(\bar{a},\bar{b})$};
\draw[<->] (T1) -- node [midway,above ] {redefinition~\eqref{eq:Xpmredef}}  (T2);
\draw[<->] (T1) -- node [midway,right ] {T~duality}  (T1d);
\draw[<->] (T2) -- node [midway,left ] {T~duality}  (T2d);
\draw[<->,dotted] (T1d) --  (T2d);
\end{tikzpicture}
\end{equation}
where all arrows can be traversed oppositely as well of course. Clearly, $\widetilde{\mathcal{T}}(a,b)$ and $\widetilde{\mathcal{T}}(\bar{a},\bar{b})$ are related by a T~duality in $\tilde{X}^-$, followed by the coordinate redefinition \eqref{eq:Xpmredef}, followed by another T~duality in $X^-$.
If we specialise this to the case corresponding to a $T\overline{T}$ transformation \textit{only}, \textit{i.e.}\ $\delta b=0$ and $b=\bar{b}=1/2$, the transformation \eqref{eq:Xpmredef} is simply a shift,
\begin{equation}
\label{eq:simpleshift}
    X^+\to Y^+=X^+ + 2\delta a\,X^-\,,\qquad X^-\to Y^-=X^-\,.
\end{equation}
Hence the diagram above yields precisely a T-duality--shift--T-duality (TsT) sequence:
\begin{equation}
\label{eq:TsTforstaticTTBar}
\begin{tikzpicture}[baseline=(current  bounding  box.center)]
\node[align=center] (T1) at (-4cm,1cm) {${\mathcal{T}}(a)$\\$(X^+,X^-)$};
\node[align=center] (T1d) at (-4cm,-1cm) {$\widetilde{\mathcal{T}}(a)$\\$(X^+,\widetilde{X}^-)$};
\node[align=center] (T2) at (4cm,1cm) {${\mathcal{T}}(\bar{a})$\\$(Y^+,Y^-)$};
\node[align=center] (T2d) at (4cm,-1cm) {$\widetilde{\mathcal{T}}(\bar{a})$\\$(Y^+,\widetilde{Y}^-)$};
\draw[<->] (T1) -- node [midway,above ] {shift~\eqref{eq:simpleshift}}  (T2);
\draw[<->] (T1) -- node [midway,right ] {T~duality}  (T1d);
\draw[<->] (T2) -- node [midway,left ] {T~duality}  (T2d);
\draw[<->] (T1d) -- node [midway,above ] {TsT}  (T2d);
\end{tikzpicture}
\end{equation}

As we remarked, changing the light-cone gauge parameters while keeping the string length fixed---a $T\overline{T}$ deformation---results in a change of the original background that is rather subtle, as it affects the \textit{global} aspects of the geometry. However, things simplify considerably by T~dualising and viewing the $T\bar{T}$ deformation as a TsT transformation. In the TsT picture, the deformation is a true deformation of the metric, and cannot be removed by a diffeomorphism (at least in general). This gives us a family of backgrounds, which in static gauge manifestly give us a Lagrangian density equal to the $T\bar{T}$ deformation of the original light-cone gauge fixed string. If we treat the parameter in this family of backgrounds as a gauge parameter, i.e. we also vary the string length ($P_- = P_-(a)$), we do nothing. In the dual picture, we would have to adjust the periodicity conditions of $\widetilde{X}^-$, because here $R$ is related to the range of $\widetilde{X}^-$, and momentum becomes winding:
\begin{equation}
\begin{tikzpicture}[baseline=(current  bounding  box.center)]
\node[align=center, rounded corners, draw=gray] (T1) at (-3cm,0cm) {${\mathcal{T}}(a)$\\$X^+=\tau,\quad p_-=2,$\\
$2R=\int_0^R\de\sigma\,p_-$};
\node[align=center, rounded corners, draw=gray] (T2) at (3cm,0cm) {${\widetilde{\mathcal{T}}}(a)$\\$X^+=\tau,\quad\widetilde{X}^- =2\sigma,$\\
$2R=\int_0^R\de\sigma\,\partial_\sigma\widetilde{X}_-$};
\node[align=center] at (0,0) {$\Longleftrightarrow$};
\end{tikzpicture}
\end{equation}
This is in agreement with the fact that a TsT transformation can be undone by a twist of the boundary conditions of the coordinates involved~\cite{Frolov:2005dj}, and in line with our expectation that only \textit{global} features of the geometry are affected. Here, the nontrivial metric deformation is exactly what we want to keep. In other words, doing a $T\overline{T}$ deformation instead of a gauge transformation from the T-dual perspective amounts to redefining the metric without keeping track of any twist of the boundary conditions. Hence the TsT approach makes more manifest the \textit{geometrical} effect of a $T\overline{T}$ deformation.

\subsection{TsT and boundary conditions}

As we mentioned,  it is well established that a TsT transformation of a sigma model is classically equivalent to twisting the boundary conditions of the sigma model before the TsT transformation \cite{Frolov:2005dj,Alday:2005ww,vanTongeren:2018vpb}. These twisted boundary conditions affect the fields associated with the TsT transformation, in our case $X^+$ and $X^-$. Concretely a TsT transformation of the type~\eqref{eq:TsTforstaticTTBar} corresponds to the boundary conditions
\begin{equation}
\begin{aligned}
    Y^+(R) - Y^+(0) & = X^+(R) - X^+(0) + 2 \delta a\, \tilde{P}_-,\\
    \widetilde{Y}^-(R) - \widetilde{Y}^-(0) & = \widetilde{X}^-(R) - \widetilde{X}^-(0) - 2 \delta a\, P_+.
\end{aligned}
\end{equation}
Such a twist of the boundary conditions can usually be equivalently viewed as a Drinfel'd-Reshetikhin twist~\cite{Drinfeld:1989st,Reshetikhin:1990ep} of the S~matrix, of the form
\begin{equation}
\label{eq:drinfeld}
    e^{i \gamma\, \epsilon^{kl}\hat{Q}_k \otimes \hat{Q}_{l}}\,,
\end{equation}
for some $\gamma\in\mathbb{R}$ and depending on the Cartans~$\hat{Q}_j$ relative to the twisted coordinates. This picture, and the effect of this twist, is quite clear when such Cartans act linearly on the particleswe of the theory; in the simplest case, they correspond to the particle flavours, and $\hat{Q}_j$ is proportional to the number operator for a given particle flavour. In our case the situation is not as transparent, because the charges corresponding to $P_{+}$ and $\widetilde{P}_-$ are not number operators in the Fock space. In general, the charges corresponding to the longitudinal isometries may not be linearly realised on the Fock space. However for our particular gauge choice, both $P_{+}$ and $\widetilde{P}_-$ act diagonally on a single-particle state. To evaluate the value of $P_{+}$ and $\widetilde{P}_-$ on a one-particle state of momentum~$p_j$ we have to recall the static-gauge fixing, which for $b=1/2$ takes the form $X^+=\tau$, $\tilde{X}^-=2\sigma$. Then as we have seen in eqs.~\eqref{eq:pplus} and~\eqref{eq:ptilde} we have that $H_{\text{w.s.}}=-P_+$ and $P_{\text{w.s.}}=2P_-$, so that
\begin{equation}
  P_+(p_j)= -\omega_j(p_j)\,,\qquad \widetilde{P}_-(p_j)=\frac{1}{2}p_j\,.
\end{equation}
Based on this, we expect the S-matrix to undergo a Drinfel'd-Reshetikhin twist of the form~\eqref{eq:drinfeld}. Considering for simplicity an  S~matrix of the form~\eqref{eq:smat} such a twist would yield
\begin{equation}
\begin{aligned}
        &S_{i_j i_k}^{i_k i_j}(p_j,p_k)\to
        S_{i_j i_k}^{i_k i_j}(p_j,p_k;\delta a)&=&e^{2i \delta a [\widetilde{P}_{+}(p_j)P_{-}(p_k)-P_{-}(p_j)\widetilde{P}_{+}(p_k)]}\,S_{i_j i_k}^{i_k i_j}(p_j,p_k)\\
        &&=&e^{i \delta a [p_j\omega_{i_k}(p_k)-p_k\omega_{i_j}(p_j)]}\,S_{i_j i_k}^{i_k i_j}(p_j,p_k)\,.
\end{aligned}
\end{equation}
We see that this precisely matches the CDD factor~\eqref{eq:cdd}.
Below we will illustrate these ideas on some examples.

\section{First example: pp-wave geometries}
\label{sec:example}
Let us consider a pp-wave metric
\begin{equation}
\label{eq:ppwavemetric}
    ds^2 = 4 \de X^+\de X^-  - V(X^i)\, \de X^+\de X^+ +\de X^i\de X^i\,.
\end{equation}
%[ST: I had to change the sign on $G_{+-}$ to make this work nicely with our conventions and flat space for $V=0$. It is of course just the diffeo $X^-\rightarrow - X^-$.]
We will consider the case where the theory has a quadratic action and is hence solvable, which is the case when
\begin{equation}
    V=\text{const.}\,,\qquad\text{or}\qquad V(X^i)= \sum_{i} (\mu_i X^i)^2\,.
\end{equation}
In practice we could complete this to a supersymmetric model, as well as possibly include a non-trivial B-field with $H=dB = C_{ij} \de X^+\wedge \de X^i\wedge \de X^j$,%
\footnote{Such a B-field plays an important role in particular in  $\AdS{3}/\CFT{2}$~\cite{Berenstein:2002jq,Russo:2002rq,Dei:2018yth} where they allow for a particularly simple exact S~matrix~\cite{Hoare:2013pma,Baggio:2018gct, Dei:2018mfl,Dei:2018jyj}.}
but we will refrain doing so to avoid cluttering our analysis. In fact, our analysis will be perhaps most interesting in the simplest case $V(X^i)=\text{const.}$, \textit{i.e.}\ for a flat spacetime.

% LET US LEAVE THIS OUT FOR NOW
%This can also be supplemented by a $B$-field such that
%\begin{equation}
%    H=dB = C_{ij} \de X^+\wedge \de X^i\wedge \de X^j\,,
%\end{equation}
%where $C_{ij}$ is a constant tensor (independent of $X^i$). This case is relevant for $\AdS_3/\CFT_2$~\cite{??}.

\paragraph{Shift of the light-cone coordinates.}
We can consider changing the gauge parameters $a\to \bar{a}=a-\delta a$ and $b\to \bar{b}=b-\delta b$ introduced above. This changes the form of the light-cone components of the metric. It is insightful to consider two simple cases. Let us first consider changing $b\rightarrow b- \delta b$. In terms of the new light-cone coordinates, the original metric now gives light cone components
\begin{equation}
G_{+-}=4+\delta b\,\frac{4(1-2a)}{\Delta_{ab}},\quad
G_{++}=-V+\delta b\, \frac{4}{\Delta_{ab}},\quad
G_{--}=0.
\end{equation}
We can see that, up to rescaling $X^+$, we have a simple change of the potential $V(X^i)$. The most interesting case,  and the one related to $T\overline{T}$ deformations, is changing $a\rightarrow a - \delta a$, which we do for simplicity at $b=1/2$. This gives
\begin{equation}
\label{eq:shiftedmetric}
\begin{aligned}
&G_{+-}=2-2\delta a\, V,\quad
&&G_{++}=-V,\quad
&&G_{--}=4 \delta a \,(2-\delta a\, V),\\
&G^{+-}=\frac{1-\delta a\, V}{2},\quad
&&G^{++}=-\delta a(2-\delta a V),\quad
&&G^{--}=\frac{V}{4},
\end{aligned}
\end{equation}
where we suppressed the $X^i$-dependence in $V$.

\subsection{Hamiltonian and spectrum of the deformed theories}
Let us now fix light-cone gauge with $X^+=\tau$ and $p_-=2$ (for $b=1/2$). The Hamiltonian can be easily found from the Virasoro constraints~\cite{Arutyunov:2009ga}
\begin{equation}
\label{eq:messyH}
\begin{aligned}
    &-p_+=\Big[\Big(1+2\delta a(p_ip_i+\acute{X}^i\acute{X}^i)+\delta a^2(16(\acute{X}^-)^2-(p_ip_i+\acute{X}^i\acute{X}^i)V)\\
    &\qquad\quad-16\delta a^3 (\acute{X}^-)^2V +4\delta a^4(\acute{X}^-)^2V^2\Big)^{1/2}-(1+\delta a V)\Big]\times\Big[\delta a (2-\delta a\, V)\Big]^{-1}\,,
\end{aligned}
\end{equation}
where $\acute{X}^-=-p_i\acute{X}^i/2$.
This is not a particularly transparent equation. However, expanding in the deformation parameter we recover
\begin{equation}
\begin{aligned}
    &-p_+= \frac{1}{2}p_ip_i+\frac{1}{2}\acute{X}^i\acute{X}^i+\frac{1}{2}V(X^i)\\
    &\qquad\qquad-\frac{\delta a}{4}\Big[(p_ip_i+\acute{X}^i\acute{X}^i+4\acute{X}^-)(p_ip_i+\acute{X}^i\acute{X}^i-4\acute{X}^-)-V(X^i)^2\Big]+O(\delta a^2),
\end{aligned}
\end{equation}
which is the free pp-wave Hamiltonian at $\delta a=0$, corrected by quartic interaction terms a leading order in~$\delta a$.

\subsection{Spectrum of the deformed theory}
\label{sec:spectrum}
The spectrum of the deformed theory can be found in principle from the Hamiltonian~\eqref{eq:messyH}. However, it is simplest to derive this from the form of the deformed S-matrix. The undeformed theory at $\delta a=0$ is free. The dispersion relation is
\begin{equation}
    \omega_i(p)=\sqrt{c^2p^2 + \mu_i^2}\,,
\end{equation}
where $c$ depends on the string tension, and the S-matrix is the identity. Hence the spectrum, for $b=1/2$ and $a=0$, is fixed by the quantisation condition
\begin{equation}
    1=e^{i p_j R}=e^{i p_j J} \qquad\Rightarrow\quad p_j=\frac{2\pi n_j}{J}\,,\qquad j=1,\dots M\,,
\end{equation}
subject to the level-matching constraint $\sum_j n_j=0$ so that
\begin{equation}
    H_{\text{w.s.}}=E-J=\sum_{j=1}^M \omega_{i_j}(\tfrac{2\pi}{J}n_j)\,.
\end{equation}
If we consider the deformed theory we have that the quantisation condition is modified by
\begin{equation}
    1=e^{i p_j (R+\delta a\,H_{\text{w.s.}})} \qquad\Rightarrow\quad p_j=\frac{2\pi n_j}{J+\delta a\,H_{\text{w.s.}}}\,,
\end{equation}
so that for the energy we have
\begin{equation}
    H_{\text{w.s.}}=E-J=\sum_{j=1}^M \omega_{i_j}\Big(\frac{2\pi\,n_j}{J+\delta a\,H_{\text{w.s.}}}\Big)\,.
\end{equation}

\paragraph{The case of flat space.}
The above equation cannot be solved in closed form unless $\mu_i=0$, which is the flat-space case. In that case we have $\omega(p)=c|p|$, so that we can introduce left- and right-movers with
\begin{equation}
    N = \sum_{i:n_i>0} n_i\,,\qquad
    \widetilde{N} = -\sum_{i:n_i<0} n_i\,.
\end{equation}
Hence we get the familiar equation
\begin{equation}
    H_{\text{w.s.}}= \frac{4\pi\,c}{J -\delta a\,H_{\text{w.s.}}}\,,\qquad
    H_{\text{w.s.}}=E-J=\frac{\sqrt{J^2+16\pi\,c\,\delta a\, N}-J}{2\delta a}\,,
\end{equation}
where we used that $N=\widetilde{N}$. We recover the fact that going from $a=0$ to $a=1/2$, with $\delta a=1/2$, sends us from the free pp-wave geometry $\de s^2=4\de X^+\de X^- +\de X^i \de X^i$ to the flat-space one, where indeed
\begin{equation}
    E= \sqrt{J^2+ 8\pi\,c\,N}\,.
\end{equation}

\begin{center}
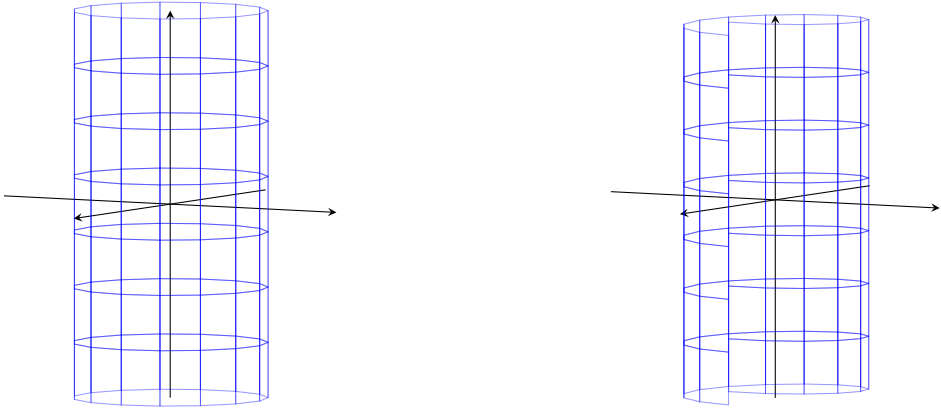
\begin{figure}[t]
\begin{tikzpicture}
\begin{axis}[xshift= -4cm,
            axis lines=center,
            ticks=none,
            xrange=-2:2,
            yrange=-2:2,
            unit vector ratio=1 1 1,
            scale=1,view/h=120, view/v=5]
    \addplot3[%
    opacity = 0.4,
    mesh,
    blue,
    z buffer = sort,
    samples = 16,
    samples y=8,
    variable = \u,
    variable y = \v,
    domain = 0:360,
    y domain = -1:1,
    ]
    ({0.25*cos(u)}, {0.25*sin(u)}, {0.5*v});
\end{axis}
\begin{axis}[xshift= 4cm,
            axis lines=center,
            ticks=none,
            xrange=-2:2,
            yrange=-2:2,
            unit vector ratio=1 1 1,
            scale=1,view/h=120, view/v=5]
    \addplot3[%
    opacity = 0.4,
    mesh,
    blue,
    z buffer = sort,
    samples = 16,
    samples y=8,
    variable = \u,
    variable y = \v,
    domain = 0:360,
    y domain = -1:1,
    ]
    ({0.25*cos(u)}, {0.25*sin(u)}, {0.5*v-0.0001*u});
\end{axis}
\end{tikzpicture}
\caption{The embedding of $(Y^+,Y^-)$ in $\mathbb{R}^{1,2}$ before and after the shift. This submanifold corresponds to the target space geometry; in the static gauge $Y^+\sim \tau$ and $Y^-\sim \sigma$ the string worldsheet has the same topology. Left: before the shift eq.~\eqref{eq:embeddingnormal} has periodic boundary conditions. Right: after the shift eq.~\eqref{eq:embeddingtwist} has twisted boundary condtions proportional to $\delta a$.}
\label{fig:twist}
\end{figure}
\end{center}

\subsection{Geometric interpretation of the shift}
We have seen that a transformation with $\delta a=1/2$ sends us from a metric of the form
\begin{equation}
\label{eq:twistedcylinder}
    \de s^2=-\de X^+\de X^+ + 2\de X^+\de X^- +\de X^i \de X^i
\end{equation}
to one of the form
\begin{equation}
\label{eq:flatcylinder}
    \de s^2= -\de Y^+\de Y^+ + \de Y^-\de Y^- +\de X^i \de X^i\,.
\end{equation}
Both these metrics define flat spaces, yet the string spectra are substantially different. This is because the two resulting manifolds, despite being \textit{locally} isomorphic, are \textit{globally} different unless we define non-trivial boundary conditions for the metric~\eqref{eq:twistedcylinder}. In eq.~\eqref{eq:flatcylinder} $Y^+$ is the time coordinate, with range~$\mathbb{R}$, while $Y^-$ is a space coordinate with some \textit{e.g.}\ range $2\pi R_Y$. The whole cylinder can be embedded in $\mathbb{R}^{1,2}\ni(t,z_1,z_2)$ as
\begin{equation}
    \label{eq:embeddingnormal}
    (t, z_1,z_2) = \big(Y^+, \cos\tfrac{Y^-}{R_Y},\sin\tfrac{Y^-}{R_Y}\big)\,,
\end{equation}
Under a true diffeomorphism we would have a different embedding
\begin{equation}
    \label{eq:embeddingtwist}
    (t, z_1,z_2) = \big(Y^+-2\delta a Y^-, \cos\tfrac{Y^-}{R_Y},\sin\tfrac{Y^-}{R_Y}\big)\,,
\end{equation}
We can conclude that the linear transformation $Y^+= X^+-2\delta a\,X^-$ which relates \eqref{eq:flatcylinder} to \eqref{eq:twistedcylinder} is not a diffeomorphism unless we correctly keep track of the boundary conditions of the fields, see figure~\ref{fig:twist}. The difference will become even more transparent in static gauge, as we shall see in the next section.

\subsection{TsT-deformed geometry}
If the take the view that a deformation $a\to\bar{a}=a-\delta a$ should be seen from the static gauge, then the background undergoes a TsT transformation. Starting from the geometry \eqref{eq:ppwavemetric}, we would like to T~dualize in $X^-$. This however is problematic since $X^-$ is null. Fortunately this problem disappears for any other member of our family of deformed backgrounds. Put differently, we want to consider the TsT transformation of a T~dual of a background, but since two of the T~dualities cancel out,  we are really just considering an ``sT'' transformation, and after the shift we no longer have issues with null coordinates. Indeed, if we shift our coordinates as in \eqref{eq:simpleshift} we obtain
\begin{equation}
\label{eq:shiftedppwavemetric}
    ds^2 = 4(1-\delta a\, V) \de Y^+\de Y^-  - V\, \de Y^+\de Y^+ + 4
    \delta a \,(2-\delta a\, V) \de Y^-\de Y^- +\de X^i\de X^i\,.
\end{equation}
As long as $\delta a$ is nonzero, $Y^-$ is not null. T~dualising in $Y^-$ now gives
\begin{equation}
\label{eq:TsTofTdualofPlanewave}
    \begin{aligned}
        ds^2 & = \frac{-4 \de Y^+ \de Y^+ + \de\widetilde{Y}^- \de\widetilde{Y}^-}{4 \delta a\,(2-\delta a\, V)} + \de X^i \de X^i,\\
        B & = - \frac{1}{\delta a} \frac{1-\delta a\, V}{2-\delta a \,V}\, \de Y^+ \wedge \de\widetilde{Y}^-.
    \end{aligned}
\end{equation}
This is our TsT transformed background.\footnote{Put differently, if we TsT transform this, the first T~duality takes us back to \eqref{eq:shiftedppwavemetric}, the shift then amounts to changing the value of $\delta a$, and the second T~duality brings us back to the above background \eqref{eq:TsTofTdualofPlanewave} with a different value of $\delta a$. In other words, for generic $\delta a$ eq. \eqref{eq:TsTofTdualofPlanewave} gives the TsT transformation of the T-dual geometry of the plane wave. It just happens to degenerate at $\delta a = 0$, the point of would-be null T~duality.} The problem in the geometry at $\delta a =0$  reflects our inability to T~dualize in a null direction. Taking this geometry and fixing a static gauge, by definition gives the gauge fixed Hamiltonian density of \eqref{eq:messyH}, which is nevertheless finite (and free) at $\delta a = 0$. In the flat-space case, where $V=\text{const.}$, we get the flat Minkowski metric with an overall scale in front of $Y^+$, $\widetilde{Y}^-$ and a constant B-field. Once again this affects the spectrum when we impose the static gauge conditions.

\section{Second example: Lin-Lunin-Maldacena geometries}
\label{sec:LLM}

One of the reasons to consider $T\overline{T}$ deformations is to construct new integrable models starting from known ones. In the context of string sigma models, the AdS$_5\times$S$^5$ type IIB superstring ~\cite{Metsaev:1998it,Bena:2003wd} is a prime example to consider deforming. At the same time, our methods are not restricted to integrable models. As a second illustrative example, let us therefore consider a more general, not generically integrable, class of string backgrounds containing AdS$_5\times$S$^5$, where the $T\overline{T}$ deformation can be neatly accounted for: Lin-Lunin-Maldacena (LLM) geometries~\cite{Lin:2004nb}.

\subsection{Some essential facts about LLM geometries}

The geometries constructed in ref.~\cite{Lin:2004nb} manifestly preserve a $\so(4)\oplus\so(4)\oplus\u(1)$ bosonic algebra. Furthermore, they are required to preserve half of the maximal amount of supercharges, \textit{i.e.}\ 16 real supercharges. These assumptions result in an ansatz for the whole supergeometry~\cite{Lin:2004nb}, where the line element is
\begin{equation}
\label{eq:LLMgeneral}
    \de s^2=
    -y(e^G+e^{-G})\big(\de t+V_i\de x^i\big)^2+\frac{\de y^2+\de x^i\de x^i}{y(e^G+e^{-G})}+
    y e^G \de \Omega_3{}^2+
    y e^{-G} \de \Omega_3'{}^2\,,
\end{equation}
where the potential $V_1(y,x_1,x_2)$, $V_2(y,x_1,x_2)$ as well as the function $G(y,x_1,x_2)$ are fixed in terms of a single function $z(y,x_1,x_2)$:
\begin{equation}
\label{eq:LLMVeq}
    z = \frac{1}{2}\frac{e^{2G}-1}{e^{2G}+1}\,,\qquad
    y\partial_y V_i=\epsilon_{ij}\partial_jz\,,\qquad
    y(\partial_iV_j-\partial_jV_i)=\epsilon_{ij}\partial_yz\,.
\end{equation}
Moreover, the $y$-dependence in $z(y,x_i)$ is fixed by a Laplace-like equation and that on the plane $y=0$ the function is piecewise constant, $z(0,x_i)=\pm\tfrac{1}{2}$. Using this, it is possible to consider a vast class of geometries, including pp-wave ones.

\paragraph{Geometries with additional rotation symmetry.} For our purposes it is convenient to restrict ourselves to geometries that possess one further $\u(1)$ isometry, corresponding to rotations in the $(x_1,x_2)$ plane. Calling $(r,\varphi)$ the radial and angular coordinates in that plane, the metric~\eqref{eq:LLMgeneral} simplifies and
\begin{equation}
\label{eq:LLMu1}
    \de s^2=
    -y(e^G+e^{-G})\big(\de t+V_\varphi\de \varphi\big)^2+\frac{\de y^2+\de r^2+r^2\de\varphi^2}{y(e^G+e^{-G})}+
    y e^G \de \Omega_3{}^2+
    y e^{-G} \de \Omega_3'{}^2\,,
\end{equation}
and now $G$ and $V_\varphi=-r\sin\varphi\, V_1+ r\cos\varphi\, V_2$ depend only on $(y,r)$. Furthermore, on the $y=0$ plane $z(0,r)$ is given by rings where values of $z=\pm\tfrac{1}{2}$ alternate. The general solution for $z(y,r)$ is then~\cite{Lin:2004nb}
\begin{equation}
    z(y,r) = \frac{(-1)^{M}}{2}+\sum_{i=0}^M (-1)^{i+1}\,\zeta(y,r; r_i)\,,
\end{equation}
with
\begin{equation}
    \zeta(y,r; r_i)= \frac{1}{2}\left(\frac{r^2-r_i^2+y^2}{\sqrt{(r^2+r_i^2+y^2)^2-4r_i^2r^2}}-1\right)\,.
\end{equation}
Indeed $\zeta(0,r;r_i)=(\text{sgn}[r^2-r_i^2]-1)/2$, so that $z(0,r)$ asymptotes to $(-1)^{M}$ at large $r$ and is always $-1/2$ at $r=0$.%
\footnote{This is a slightly different normalisation with respect to ref.~\cite{Lin:2004nb}, as we will be interested in changing the large-$r$ behaviour later on.}
We can also solve the equation~\eqref{eq:LLMVeq} for $V_\varphi$ to find
\begin{equation}
    V_\varphi(y,r) = \psi_\varphi(r)+\sum_{i=1}^M (-1)^{i+1}\,v(y,r; r_i)\,,\end{equation}
    with
    \begin{equation}
    v(y,r; r_i) = -\frac{1}{2}\left(\frac{r^2+y^2+r_i^2}{\sqrt{(r^2+y^2+r_i^2)^2-4r_i^2r^2}}-1\right)\,.
\end{equation}
This solution differs from the one in ref.~\cite{Lin:2004nb} by the function~$\psi_\varphi(r)$ which, looking back at~\eqref{eq:LLMVeq}, must be $y$-independent and should yield an irrotational vector field $(\psi_1,\psi_2)$ in the $(x_1,x_2)$ plane. If we require $V_\varphi$ to be well-defined at $r=0$ and $r=\infty$, it must be that $\psi_\varphi(r)=0$.

\paragraph{Undeformed AdS$_5\times$S$^5$.} Among the many LLM geometries, we can recover undeformed AdS$_5\times$S$^5$ by simply setting $M=0$, $\psi(r)=0$, and performing the change of variables~\cite{Lin:2004nb}
\begin{equation}
\label{eq:LLMredef}
    y= r_0\, \sin\theta\,\sinh\rho\,,\qquad
    r = r_0\,\cos\theta\,\cosh\rho\,\qquad
    \varphi=\phi-t\,.
\end{equation}
This gives the line element of AdS$_5\times$S$^5$ in global coordinates
\begin{equation}
\label{eq:adsmetric}
    \de s^2=
    r_0\Big[
    -\cosh^2\rho\,\de t^2+\de\rho^2+\sinh^2\rho\,\de\Omega_3{}^2
    +\cos^2\theta\,\de \phi^2+\de\theta^2+\sin^2\theta\,\de\Omega_3'{}^2
    \Big]\,.
\end{equation}

\subsection{Deforming the LLM geometries}
It is natural to ask whether the deformation discussed above can be applied to an LLM geometry to obtain a geometry of the same type.
We may address this question in the direct geometry or in the T-dual one. Here it is most illustrative to work in terms of the direct geometry, where we consider the shift~\eqref{eq:simpleshift}.
The shift deformation makes sense in the case where we have an $\u(1)^{\oplus 2}$ symmetry on top of the $\so(4)^{\oplus 2}$, because (a combination of) the two $\u(1)$ directions will play the role of the shift symmetries $X^\pm$ appearing in the light-cone gauge fixing. Moreover, by construction, the shift deformation preserves the full $\so(4)^{\oplus2}\oplus\u(1)^{\oplus2}$ symmetry. For AdS$_5\times$S$^5$, it clearly will also preserve the $\su(2|2)^{\oplus 2}$ (centrally extended) symmetry which is manifest after gauge fixing~\cite{Arutyunov:2009ga}. It is actually relatively straightforward to reverse-engineer what the shift of section~\ref{sec:deformed-bgd} is in the LLM language. Since the shift does not affect the angular part of the line element, it is reasonable to look for a transformation affecting $V_\varphi$ only. Consider the redefinition
\begin{equation}
    V_\varphi(y,x_1,x_2) \mapsto V_\varphi(y,x_1,x_2) + \alpha\,.
\end{equation}
In Cartesian components this amounts to $V_i\mapsto V_i+ \alpha\,\psi_i$ with $ \psi_i=\epsilon_{ij}\partial_j \log r$. This is clearly irrotational wherever it is defined, and yields a new solution of the LLM constraints.
To compare with the shift transformation discussed in eq.~\eqref{eq:simpleshift} it is convenient to introduce light-cone coordinates. As evidenced by eq.~\eqref{eq:LLMredef}, $\phi$ is already a light-cone coordinate, and in our notation of eq.~\eqref{eq:simplelc}, $\varphi=2 X^-$ while $t=X^+ - X^-$. Hence the line element~\eqref{eq:LLMu1} becomes
\begin{equation}
    \de s^2=
    -y(e^G+e^{-G})\big(\de X^+ +(2V_\varphi -1)\de X^-\big)^2+\frac{\de y^2+\de r^2+4r^2(\de X^{-})^2}{y(e^G+e^{-G})}+
    \dots\,,
\end{equation}
where the ellipsis denote the angular part of the line element, which is unchanged.
We can see that the modification
\begin{equation}
    V_\varphi\mapsto V_\varphi+\alpha \quad \text{is equivalent to}\quad
    X^+ \mapsto X^+ + 2\delta a\, X^- \quad\text{for}\quad \alpha=\delta a\,,
\end{equation}
while leaving $X^-$ unchanged. This is precisely the deformation of eq.~\eqref{eq:simpleshift}. This is completely general, holding for any LLM geometry with an additional $\u(1)$ symmetry.

\subsection{Speculations on gauge-theory duals}
\label{sec:speculations}

When we consider the $T\overline{T}$ deformation of strings with an AdS/CFT interpretation, it becomes natural to ask whether this deformation has a counterpart in the dual field theory. Our prescription does not say much about how to construct a hypothetical holographic dual of a deformed background, but in the case of AdS$_5\times$S$^5$ it might prove interesting to speculate a little, based on recent investigations by Caetano, Peelaers and Rastelli~\cite{Rastelli1, Rastelli2}. These authors are considering irrelevant deformations of $\mathcal{N}=4$ supersymmetric Yang-Mills (SYM), revisiting an older proposal of Intriligator~\cite{Intriligator:1999ai}. Their starting point is to consider $\mathcal{N}=4$ SYM on $\mathbb{R}\times\text{S}^3$ and to look for the ``least irrelevant'' deformation that preserve ``as much (super)symmetry'' as possible. As the deformation is irrelevant, the best that can be hoped for is to preserve the supersymmetry generators (as opposed to the superconformal ones), as well as $\so(4)^{\oplus 2}\oplus \u(1)^{\oplus 2}$. This can be done by deforming the action by a dimension-8 operator inspired by the one of considered in~\cite{Intriligator:1999ai} for SYM in $\mathbb{R}^{1,3}$. Working on on $\mathbb{R}\times\text{S}^3$ however brings several new features, as well as some technical complications~\cite{Rastelli1, Rastelli2}. The upshot appears to be that the preserved symmetries after the deformation take the form of two copies of the centrally extended $\su(2|2)$ of Beisert~\cite{Beisert:2005tm}, though the algebra is twisted with respect to the usual AdS$_5\times$S$^5$ setup.%
\footnote{Ordinarily, the light-cone Hamiltonian would emerge from the anti-commutator of supercharges and superconformal charges, and the real form of the $\psu(2,2|4)$ algebra is chosen in such a way that the two sets of supercharges are Hermitian conjugate to each other, see \textit{e.g.}\ ref.~\cite{Arutyunov:2009ga} for a review.}
This guarantees that, when setting up the spectral problem in terms of a spin chain, there is a two-to-two scattering matrix for magnons which differs from Beisert's by (at most) a CDD factor, and in particular means that the two body S~matrix satisfies the Yang-Baxter equation.\footnote{These observations are not sufficient by themselves to guarantee integrability of the deformed theory. One would need to investigate multi-magnon states to ascertain this fact; this investigation is ongoing. We thank L.~Rastelli for discussions on this point.} The $T\overline{T}$ deformation of AdS$_5\times$S$^5$ shares these general properties. Furthermore, the form of the deforming irrelevant operator in SYM bears some resemblance with the $T\overline{T}$ operator, though we are now in four dimensions. While (infinitely) many CDD factors are possible, one might wonder whether the CDD factor of the deformed SYM spin chain S matrix is related to or precisely of the $T\overline{T}$ type. Checking this requires presently unavailable explicit results in the deformed field theory.

Interestingly, Caetano, Peelaers and Rastelli have speculated that the Intriligator deformation on $\mathbb{R}\times\text{S}^3$ might be dual to a particular LLM geometry~\cite{Rastelli2}. This geometry is of the form discussed above, with $M=1$. in the LLM language it is given by a ``droplet'' in the $(x_1,x_2)$ plane for $r\leq r_0$ and another one for $r\geq r_1$. The undeformed limit is then $r_1\to\infty$. We can then at least try to compare this LLM geometry to our $T\overline{T}$ geometry constructed via the shift~\eqref{eq:simpleshift}. We will do this perturbatively, as this directly links to the perturbative S matrix as well.

We start by parametrising
\begin{equation}
\label{eq:gammaparam}
    r_1= r_0\,\sqrt{1+\frac{1}{2\gamma}}\,,
\end{equation}
so that the undeformed limit is $\gamma\to0$. Consider the undeformed case, and set $r_0=1$ for convenience. Then the metric takes the form~\eqref{eq:adsmetric}, which can be conveniently rewritten as
\begin{equation}
\de s^2=
-\left(\frac{1+\frac{x_ix_i}{4}}{1-\frac{x_ix_i}{4}}\right)^2 \de t^2
+\frac{\de x_i \de x_i}{\left(1-\frac{x_ix_i}{4}\right)^2} +
\left(\frac{1-\frac{x_mx_m}{4}}{1+\frac{x_mx_m}{4}}\right)^2 \de \phi^2
+\frac{\de x_m \de x_m}{\left(1+\frac{x_mx_m}{4}\right)^2}\,,
\end{equation}
where $(x_i)_{i=1,\dots4}$ are the four transverse coordinates of AdS$_5$ and $(x_m)_{m=5,\dots 8}$ are the four transverse coordinates of S$^5$. Working \textit{e.g.} in the $a=b=1/2$ gauge, it is straightforward to write down the light-cone Hamiltonian density~$\mathcal{H}_{\text{w.s.}}$ which fixes the tree-level S~matrix (see for instance ref.~\cite{Arutyunov:2009ga}). We have%
\footnote{The indices $\mu,\nu$ take values $1,\dots 8$, while $i,j=1,\dots 4$ and $m,n=5,\dots 8$.}
\begin{equation}
\begin{aligned}
    \mathcal{H}_{\text{w.s.}} =& \frac{1}{2} p_\mu p_\mu + \frac{1}{2}\acute{x}_\mu\acute{x}_\mu + \frac{1}{2}{x}_\mu{x}_\mu+ \frac{1}{4}x_ix_i\Big(p_jp_j+\acute{x}_j\acute{x}_j+2 \acute{x}_m\acute{x}_m\big)\\
    &-\frac{1}{4}x_mx_m\Big(p_np_n+\acute{x}_n\acute{x}_n+2 \acute{x}_i\acute{x}_i\big)+ \cdots\,,
\end{aligned}
\end{equation}
where the ellipsis indicate terms of order six or higher in the fields. The effect of a $T\overline{T}$ deformation is to change the quartic term by a linear expression in $\delta a$,
\begin{equation}
    \delta \mathcal{H}_{\text{w.s.}} = \frac{\delta a}{2}\,\Big((x_\mu x_\mu)^2 - (p_\mu p_\mu+\acute{x}_\mu\acute{x}_\mu)^2+4(\acute{x}_\mu p_\mu)^2
    \Big)+\cdots \,.
\end{equation}

Let us now consider the geometry with $\gamma\neq0$ (and no $T\overline{T}$ deformation). It is convenient to expand the metric in the transverse fields. To illustrate the dependence on~$\gamma$, let us write down the first few terms in the field expansion:
\begin{equation}
\begin{aligned}
&G_{ii}&& = \phantom{+}1+\frac{x_ix_i}{2}+\frac{3}{16}(x_ix_i)^2-\gamma(1+2\gamma) (x_\mu x_\mu)^2+\cdots\\
&G_{mm} &&= \phantom{+}1-\frac{x_m x_m}{2}+\frac{3}{16}(x_mx_m)^2+\gamma(1+2\gamma) (x_\mu x_\mu)^2+\cdots\\
&G_{tt} &&= -1+4\gamma -x_ix_i+2\gamma(5 x_ix_i-3 x_mx_m)+8\gamma^2(x_ix_i-2x_mx_m)+\cdots\\
&G_{\phi\phi} &&= \phantom{+}1+4\gamma -x_mx_m+2\gamma(3 x_ix_i-5 x_mx_m)+8\gamma^2(x_ix_i-2x_mx_m)+\cdots\\
&G_{t\phi} &&= \phantom{+1}-4\gamma-8\gamma(x_ix_i-x_mx_m)+\cdots\\
\end{aligned}
\end{equation}
Higher-order terms are collected in appendix~\ref{sec:appexplicit}. It is interesting to note that, at leading order, the effect of tuning  $\gamma$ on the light-cone coordinates is precisely that of a coordinate shift. Quite interestingly, even when we account for the transverse fields, there appears to be a close relationship between this deformation and the $\delta a$~shift considered above. Namely, up to and including quartic order in the transverse fields, the effect of tuning $\gamma$ on the Hamiltonian  is exactly the same as tuning~$\delta a$:
\begin{equation}
    \delta \mathcal{H}_{\text{w.s.}} = \frac{\gamma}{2}\,\Big((x_\mu x_\mu)^2 - (p_\mu p_\mu+\acute{x}_\mu\acute{x}_\mu)^2+4(\acute{x}_\mu p_\mu)^2
    \Big)+\cdots \,.
\end{equation}
While this is suggestive, it turns out that at higher order the two deformations start to differ. We have collected the expression of the sixth-order terms in appendix~\ref{sec:appexplicit}. We have not been able to perturbatively construct a canonical transformation linking the two deformations at this order. The presence of a genuine discrepancy between the two constructions is in line with observation that LLM geometries are generically non-integrable~\cite{Chervonyi:2013eja}, while by construction the $\delta a$-shifted geometry is integrable.

\section{Conclusions and outlook}
\label{sec:conclusion}

The uniform light-cone gauge formalism for string theory~\cite{Arutyunov:2004yx,Arutyunov:2005hd,Arutyunov:2006gs} allows one to readily construct $T\overline{T}$ deformations of various models~\cite{Baggio:2018gct,Baggio:2018rpv,Frolov:2019xzi,Frolov:2019nrr}. This starts by uplifting the original model to a reparametrisation-invariant model two higher dimensions, and then gauge-fixing this invariance appropriately. In this paper we asked what happens to this uplifted geometry under a $T\overline{T}$ deformation, \textit{i.e.}\ what the $T\overline{T}$ deformation of a (string) sigma model is. Operatively, we tune the would-be gauge parameter in the worldsheet Lagrangian \textit{only}, and not in the identification of conserved charges or volume~$R$ of the model. The effect of this deformation is subtle from the point of view of the original geometry for our light-cone gauge picture, but becomes more transparent when taking a T-dual point of view~\cite{Kruczenski:2004cn}, where we exchange light-cone gauge for \textit{static} gauge fixing. In the T-dual frame, a $T\overline{T}$ deformation affects the local geometry directly, taking the form of a TsT transformation.\footnote{In this paper we only discussed NSNS backgrounds explicitly, but RR fields can of course be added and TsT transformed.}
This TsT picture then also gives a natural interpretation to the $T\overline{T}$ CDD factor as a Drinfel'd-Reshetikhin twist of the S-matrix; this is particularly transparent thanks to the static-gauge identification of target-space charges with worldhseet momentum and energy. %We illustrated these ideas on a simple class of pp-wave backgrounds.
Computationally, for the purpose of generating deformed Lagrangians, this static-gauge approach is equivalent to the uniform ligth-cone gauge treatment of refs.~\cite{Baggio:2018gct,Baggio:2018rpv,Frolov:2019xzi,Frolov:2019nrr}; conceptually however, we feel that it further clarifies why $T\overline{T}$ deformations are so intimately related to gauge-fixed sigma models, and may help further uncover some of the features of this important class of deformations.
Let us remark that our discussion of $T\overline{T}$ deformations can be quite straightforwardly extended to $T\overline{J}$ and $J\overline{T}$ deformations, as well as to more general deformations along the lines of ref.~\cite{Frolov:2019nrr}. We briefly comment on this in appendix~\ref{sec:TJappendix}.

It would be interesting to extend our approach to include fermions and to consider fully-fledged supergeometries. The first steps have been taken while investigating the relation between $T\overline{T}$ and supersymmetry, as well as in ref.~\cite{Frolov:2019nrr} for more general theories. However, a complete analysis of such a setup, including the role of $\kappa$-symmetry, has not yet been performed.
It would also be interesting to extend this analysis to the non-relativistic deformations of refs.~\cite{Guica:2017lia,Bzowski:2018pcy,Nakayama:2018ujt,Chakraborty:2018vja, LeFloch:2019rut, Guica:2019vnb,Chakraborty:2019mdf}, which can indeed be understood in the framework of light-cone gauge~\cite{Frolov:2019xzi}, and further explore its relation with null dipole-deformed CFT~\cite{Alishahiha:2003ru,Guica:2017jmq}, which can indeed be understood in AdS/CFT by means of TsT transformations involving light-cone directions.

Another especially interesting case is that of integrable \textit{string} sigma models. Here, the $T\overline{T}$ CDD factor can be readily taken into account in their Bethe ansatz (and eventually thermodynamic Bethe ansatz). As we saw, in the special case of flat space, the $T\overline{T}$ deformation can trivialize the S matrix. In general, however, the S matrix will remain nontrivial, and be nontrivially modified. This is certainly the case for all integrable string backgrounds involving Ramond-Ramond fluxes, where the form of the light-cone symmetry algebra fixes the S-matrix to be non-diagonal.%
\footnote{%
The relationship between light-cone symmetry algebra and the integrable S~matrix was originally worked out for $\AdS{5}\times\text{S}^{5}$ in refs.~\cite{Arutyunov:2006ak,Arutyunov:2006yd}. 
}
Still, it would be interesting to study the corresponding deformations of (the T~duals of) such integrable backgrounds, as at least we have good control over the spectral problem.
In this paper we have considered two classes of backgrounds: pp-wave geometries, which are integrable, and LLM geometries, which are not generally integrable with the important exception of AdS$5\times$S$^5$. In both cases we derived explicit expressions for the deformed backgrounds. In particular, for AdS$5\times$S$^5$, we have described a ``shifted'' geometry which would yield a $T\overline{T}$ deformation of Beisert's S~matrix, though it is not clear what interpretation this would have for the gauge-theory dual.%
\footnote{Similar deformations are currently being investigated in $\mathcal{N}=4$ SYM~\cite{Rastelli:paper}.}

One could also study deformed AdS backgrounds in the T-dual frame, by means of a TsT transformation rather than a shift. As an illustration, for AdS$_2\times$S$^2$ in global coordinates
\begin{equation}
    ds^2 = -(1+\rho^2)dt^2 + \frac{d\rho^2}{1+\rho^2}  +(1-r^2)d\phi^2 + \frac{dr^2}{1-r^2},
\end{equation}
with isometry coordinates $t$ and $\phi$ as input for the light-cone coordinates, the dual deformed geometry takes the form
\begin{equation}
\begin{aligned}
    ds^2 & = \frac{-(1-r^2)(1+\rho^2) \de Y^+ \de Y^+ +\frac{1}{4}\de \widetilde{Y}^- \de\widetilde{Y}^-}{1-r^2 + 2 \delta a \, (1-r^2) - \delta a^2 (r^2+\rho^2)} + \frac{d\rho^2}{1+\rho^2} + \frac{dr^2}{1-r^2},\\
    B & =  -\frac{1-r^2-\delta a \, (r^2+\rho^2)}{1-r^2 + 2 \delta a \, (1-r^2) - \delta a^2 (r^2+\rho^2)} \de Y^+ \wedge \de\widetilde{Y}^-,
\end{aligned}
\end{equation}
where deform away from $a=0$.\footnote{Unlike the pp-wave example of the last section, here we generically never encounter a null direction in the T~duality. Of course we can see the problem reappear by setting $r,\rho \rightarrow 0$ and taking $\delta a = -1/2$.}
As the $T\overline{T}$ deformation preserves integrability, it would be interesting combine it with other integrable deformations of strings, such as Yang-Baxter deformations \cite{Klimcik:2008eq,Delduc:2013qra,Kawaguchi:2014qwa}. These, as a nice contrast, contain TsT transformation of the \textit{direct} (as opposed to T-dual) geometry~\cite{Osten:2016dvf}, see also \cite{Matsumoto:2014nra,vanTongeren:2015soa,Borsato:2016ose}.

Among the integrable AdS/CFT setups, the case of AdS${}_3$ backgrounds deserves a special discussion. In that case, backgrounds can be supported by a mixture of Ramond-Ramond (RR) and Neveu-Schwarz-Neveu-Schwarz (NSNS) fluxes (see ref.~\cite{Sfondrini:2014via} for a review of integrability in this setup). The kinematics depends both on the RR strength~$h$ and the NSNS strength~$k$ and takes the form~\cite{Hoare:2013lja,Lloyd:2014bsa,Borsato:2015mma}
\begin{equation}
    \omega_i (p) = \sqrt{\Big(m_i +\frac{k}{2\pi}p\Big)^2+4h^2\sin^2\Big(\frac{p}{2}\Big)}\,.
\end{equation}
When no RR fluxes are present, $h=0$ and $k\in\mathbb{N}$ is the level of the $\mathfrak{sl}(2)\oplus\mathfrak{su}(2)$ supersymmetric Wess-Zumino-Witten (WZW) model; we see from~$\omega_i(p)$ that the model is \textit{chiral}, even after gauge fixing. Moreover, it can be checked that the perturbative worldsheet S~matrix is \textit{proportional to the identity}, and takes a universal dependent on the chirality of the scattered particles (but not on the masses~$m_i$)~\cite{Hoare:2013pma,Dei:2018jyj}
\begin{equation}
    S_{ij}^{kl}(p_1, p_2)=
    \begin{cases}
    e^{-i \frac{k}{2\pi}p_1p_2}\ \delta_{i}^{l}\,\delta_{j}^{k}\qquad& p_1\quad\text{left-moving},\quad p_2\quad\text{right-moving},\\
    e^{+i \frac{k}{2\pi}p_1p_2}\ \delta_{i}^{l}\,\delta_{j}^{k}\qquad& p_2\quad\text{left-moving},\quad p_1\quad\text{right-moving},\\
    \delta_{i}^{l}\,\delta_{j}^{k}&\text{otherwise}\,.
    \end{cases}
\end{equation}
This allows to solve for the spectrum in closed form~\cite{Baggio:2018gct,Dei:2018mfl,Dei:2018jyj} along the lines of what we did from flat space in section~\ref{sec:spectrum}. However, the scattering \textit{cannot} be completely trivialised by a $T\overline{T}$ deformation.%
\footnote{%
This is because in this case $p_1\omega(p_2)-p_2\omega(p_1)\neq \pm 2p_1p_2$, nor does it vanish for same-chirality scattering---a fact that is crucial in order to reproduce the spectrally-flowed sectors of the WZW description, see refs.~\cite{Dei:2018mfl,Dei:2018jyj}.}
On the other hand, for this theory it also possible to consider a $T\overline{T}$ deformation of the \textit{dual} conformal field theory. It was proposed~\cite{Giveon:2017nie,Giveon:2017myj} that these too can be studied \textit{on the worldsheet}, namely that a $T\overline{T}$~deformation on the boundary should correspond to a $J\overline{J}$~deformation on the worldsheet (which can be then analysed by worldsheet-CFT tools).Quite interestingly, such $J\overline{J}$~deformations \textit{can also be understood as TsT transformations}~\cite{Borsato:2018spz}.
This scenario can be generalised to non-relativistic $J\overline{T}$ deformations, and in that case too deformations of the dual CFT${}_2$ can be interpreted as TsT transformations~\cite{Apolo:2018qpq,Apolo:2019yfj}.%
\footnote{See appendix~\ref{sec:TJappendix} for a discussion of $J\overline{T}$ deformations on the worlsdheet of th gauge-fixed theory.}
This points to the fact that in pure-NSNS AdS${}_3$/CFT${}_2$, a rich interplay arises between deformations on the worldsheet and in the two-dimensional dual, which is yet to be explored. 
We hope to revisit some of these questions in the near future.

\section*{Acknowledgements}

We would like to thank Sergey Frolov, Ben Hoare, Leonardo Rastelli and Fiona K.~Seibold for insightful discussions and Sergey Frolov  for comments on a draft of this manuscript.
A.S.\ would like to thank the participants of the workshop \textit{A Fresh Look at AdS3/CFT2} at Villa Garbald for stimulating discussions that paved the way to this research, as well as the organisers and participants of the CERN Theory Institute \textit{Exact computations in AdS/CFT} for the stimulating environment where the last phases of this work were completed, and in particular Nikolay Bobev for discussions on AdS$_5\times$S$^5$. He is also grateful to Christian Ferko, Hongliang Jiang, Sav Sethi and Gabriele Tartaglino-Mazzucchelli for ongoing collaboration on related topics. A.S.'s work is funded by ETH Career Seed Grant No.~SEED-23 19-1, as well as by the NCCR \textit{SwissMAP}, funded by the Swiss National Science Foundation.
The work of the S.T. is supported by the German Research Foundation (DFG) via the Emmy Noether program ``Exact Results in Extended Holography''. S.T. is supported by~L.T.

\appendix
\section{Explicit expression for the LLM geometry}
\label{sec:appexplicit}

We collect here some explicit expression for the LLM geometry with~$M=2$ introduced in section~\ref{sec:LLM} as a function of the parameter~$\gamma$, see eq.~\eqref{eq:gammaparam}. Up to order six in the transverse fields, and exactly in~$\gamma$, we have
\begin{equation}
\begin{aligned}
&G_{ii}&& = \phantom{+}1+\tfrac{1}{2}x_ix_i+\tfrac{3}{16}(x_ix_i)^2-\gamma(1+2\gamma) (x_\mu x_\mu)^2+ \tfrac{1}{16}(x_ix_i)^3\\
&&& \quad +\big[\big(\gamma+6\gamma^2+8\gamma^3\big) x_mx_m-\big(\tfrac{3}{2}\gamma+7\gamma^2+8\gamma^3\big) x_ix_i\big] (x_\mu x_\mu)^2+\cdots
\end{aligned}
\end{equation}
\begin{equation}
\begin{aligned}
&G_{mm} &&= \phantom{+}1-\tfrac{1}{2}x_m x_m+\tfrac{3}{16}(x_mx_m)^2+\gamma(1+2\gamma) (x_\mu x_\mu)^2-\tfrac{1}{16}(x_m x_m)^3\\
&&&\quad+\big[\big(\gamma+6\gamma^2+8\gamma^3\big) x_ix_i-\big(\tfrac{3}{2}\gamma+7\gamma^2+8\gamma^3\big) x_mx_m\big] (x_\mu x_\mu)^2+
\cdots
\end{aligned}
\end{equation}
\begin{equation}
\begin{aligned}
&G_{tt} &&= -1+4\gamma -x_ix_i+2\gamma(5 x_ix_i-3 x_mx_m)+8\gamma^2(x_ix_i-2x_mx_m)-\tfrac{1}{2}(x_ix_i)^2\\
&&&\quad+4\gamma\big[(x_mx_m)^2-3x_mx_m\,x_ix_i+3(x_ix_i)^2\big]
\\&&&\quad +
\gamma^2\big[38(x_mx_m)^2-72x_mx_m\,x_ix_i+22(x_ix_i)^2\big]
\\&&&\quad +
8\gamma^3\big[7(x_mx_m)^2-12x_mx_m\,x_ix_i+(x_ix_i)^2\big]-\tfrac{3}{16}(x_ix_i)^3
\\&&&\quad +
\gamma\big[-\tfrac{17}{8}(x_mx_m)^3+7(x_mx_m)^2x_ix_i-12x_mx_m(x_ix_i)^2+\tfrac{79}{8}(x_ix_i)^3\big]
\\&&&\quad +
\gamma^2\big[-47(x_mx_m)^3+128(x_mx_m)^2x_ix_i-138x_mx_m(x_ix_i)^2+\tfrac{55}{2}(x_ix_i)^3\big]
\\&&&\quad +
4\gamma^3\big[-45(x_mx_m)^3+135(x_mx_m)^2x_ix_i-97x_mx_m(x_ix_i)^2+3(x_ix_i)^3\big]
\\&&&\quad -
16\gamma^4\big[12(x_mx_m)^3-39(x_mx_m)^2x_ix_i+20x_mx_m(x_ix_i)^2+(x_ix_i)^3\big]
+\cdots
\end{aligned}
\end{equation}
\begin{equation}
\begin{aligned}
&G_{\phi\phi} &&= \phantom{+}1+4\gamma -x_mx_m+2\gamma(3 x_ix_i-5 x_mx_m)+8\gamma^2(x_ix_i-2x_mx_m)
\\&&&\quad
+\tfrac{1}{2}(x_mx_m)^2
+4\gamma\big[3(x_mx_m)^2-3x_mx_m\,x_ix_i+(x_ix_i)^2\big]
\\&&&\quad
+\gamma^2\big[50(x_mx_m)^2-72x_mx_m\,x_ix_i+10(x_ix_i)^2\big]
\\&&&\quad
+8\gamma^3\big[7(x_mx_m)^2-12x_mx_m\,x_ix_i+(x_ix_i)^2\big]-\tfrac{3}{16}(x_mx_m)^3
\\&&&\quad +
\gamma\big[-\tfrac{79}{8}(x_mx_m)^3+12(x_mx_m)^2x_ix_i-7x_mx_m(x_ix_i)^2+\tfrac{17}{8}(x_ix_i)^3\big]
\\&&&\quad +
\gamma^2\big[-81(x_mx_m)^3+166(x_mx_m)^2x_ix_i-92x_mx_m(x_ix_i)^2+\tfrac{3}{2}(x_ix_i)^3\big]
\\&&&\quad -
4\gamma^3\big[55(x_mx_m)^3-149(x_mx_m)^2x_ix_i+79x_mx_m(x_ix_i)^2+3(x_ix_i)^3\big]
\\&&&\quad -
16\gamma^4\big[12(x_mx_m)^3-39(x_mx_m)^2x_ix_i+20x_mx_m(x_ix_i)^2+(x_ix_i)^3\big]
+\cdots
\end{aligned}
\end{equation}
\begin{equation}
\begin{aligned}
&G_{t\phi} &&=-4\gamma-8\gamma(x_ix_i-x_mx_m)
-4\gamma\big[2(x_mx_m)^2-3x_mx_m\,x_ix_i+2(x_ix_i)^2\big]
\\&&&\quad
+\gamma^2\gamma\big[-44(x_mx_m)^2+72x_mx_m\,x_ix_i-16(x_ix_i)^2\big]
\\&&&\quad
-8\gamma^3\gamma\big[7(x_mx_m)^2+12x_mx_m\,x_ix_i+(x_ix_i)^2\big]
\\&&&\quad +
\tfrac{1}{2}\gamma\big[11(x_mx_m)^3-20(x_mx_m)^2x_ix_i+20x_mx_m(x_ix_i)^2-11(x_ix_i)^3\big]
\\&&&\quad +
\gamma^2\big[63(x_mx_m)^3-148(x_mx_m)^2x_ix_i+116x_mx_m(x_ix_i)^2-\tfrac{27}{2}(x_ix_i)^3\big]
\\&&&\quad +
\gamma^3\big[200(x_mx_m)^3-568(x_mx_m)^2x_ix_i+352x_mx_m(x_ix_i)^2\big]
\\&&&\quad +
\gamma^4\big[192(x_mx_m)^3-624(x_mx_m)^2x_ix_i+320x_mx_m(x_ix_i)^2+16(x_ix_i)^3\big]
\\&&&\quad
+\cdots
\end{aligned}
\end{equation}
where the ellipsis are of eight order or higher in the transverse fields.

Using these expressions, it is easy to work out the quadratic and quartic Hamiltonian, which are presented in section~\ref{sec:speculations}. The sixth-order Hamiltonian is considerably more involved. For $\gamma=0$ (\textit{i.e.}, for the underformed AdS$_5\times$S$^5$ background) it takes the form
\begin{equation}
\begin{aligned}
    &\mathcal{H}_{\text{w.s.}}^{(6)} = \frac{1}{32}\Big\{
    \Big[\big(p_\mu p_\mu+\acute{x}_\mu\acute{x}_\mu\big)^2  - 4 \big(p_\mu\acute{x}_\mu\big)^2- \big(x_ix_i\big)\big(x_mx_m\big)\Big](x_\mu x_\mu)\\
    &\qquad\qquad
    + p_ip_i\Big[2\big(x_mx_m\big)^2-\big(x_ix_i\big)^2\Big]+ p_mp_m\Big[2\big(x_ix_i\big)^2-\big(x_mx_m\big)^2\Big]\\
    &\qquad\qquad
    +\acute{x}_i\acute{x}_i\Big[2\big(x_mx_m\big)^2-8 \big(x_mx_m\big)\big(x_ix_i\big) +9\big(x_ix_i\big)^2\Big]\\
    &\qquad\qquad
    +\acute{x}_m\acute{x}_m\Big[2\big(x_ix_i\big)^2-8\big(x_ix_i\big) \big(x_mx_m\big) +9\big(x_mx_m\big)^2\Big]\Big\}\,.
\end{aligned}
\end{equation}
Identifying the deformation parameter $\gamma=\delta a$, like it appears from the quartic term, we can compute the difference $\Delta$ between the sixth-order $\gamma$-deformed and $\delta a$-deformed Hamiltonian. We have
\begin{equation}
\begin{aligned}
    &\Delta=\frac{\gamma(1+2\gamma)}{4}\Big\{
    3(x_ix_i-x_mx_m) \Big[\big(p_\mu p_\mu+\acute{x}_\mu\acute{x}_\mu\big)^2  - 4 \big(p_\mu\acute{x}_\mu\big)^2\Big]\\
    &\qquad
    +\Big[2\big(
    p_mp_m-p_ip_i-\acute{x}_m\acute{x}_m+\acute{x}_i\acute{x}_i
    \big)
    +x_mx_m-x_ix_i\Big]\big(x_\mu x_\mu\big)^2
    \Big\}\,.
\end{aligned}
\end{equation}

\section{\texorpdfstring{$T\overline{J}$}{TJbar}, \texorpdfstring{$J\overline{T}$}{JTbar}, and TsT}
\label{sec:TJappendix}

In the main text we discussed the geometric interpretation of $T \overline{T}$ deformations as TsT transformations in the T dual frame. It is natural to ask whether a similar interpretation exists for deformations of $T\overline{J}$ and $J\overline{T}$ type. This is indeed the case, even though only a limited number of such deformations have a simple geometric interpretation in a given T-dual frame.

In order to be able to consider generalized deformations we need to assume that our background has a further $\mathfrak{u}(1)$ isometry commuting with two light-cone isometries. Let us fix coordinates such that this extra isometry acts as a shift in $X^1$. This direction can now be mixed in to TsT transformations. In general, given $m$ commuting isometries we can consider $m(m-1)/2$ independent TsT transformations, giving us three with isometries in $X^+,\tilde{X}^-$ and $X^1$.

For concreteness let us consider a TsT transformation in $(\tilde{X}^-,X^1)$. Since we are doing a TsT transformation starting from the static-gauge frame, from the point of view of light-cone-gauge description we are doing an sT transformation, shifting $X^1 \rightarrow X^1 + \alpha X^-$, and T dualising $X^- \rightarrow \tilde{X}^-$. This shift in the original geometry is precisely what corresponds to the canonical transformation giving a $JT_{\mu}$ deformation with~$\mu=\sigma$, the spatial direction on the worldsheet. Indeed, as discussed in \cite{Frolov:2019xzi}, \textit{cf.}\ section 2.2 point 3, this canonical transformation is
\begin{equation}
    X^1 \rightarrow X^1 - a_{1-} X^-,\quad X^- \rightarrow X^-,\quad p_1 \rightarrow p_1,\quad p_- \rightarrow p_- + a_{1-} p_1.
\end{equation}
For $\alpha = -a_{1-}$ this agrees exactly with our shift; the shift in momenta follows directly from the shift of coordinates. To complete the picture we just perform one more T-duality in $\widetilde{X}^-$, which takes us back to the static-gauge picture.

In the main text we saw that a TsT in $(\tilde{X}^-,X^+)$ gives the $T \overline{T}$ deformation, and we just discussed that one in $(\tilde{X}^-,X^1)$ gives a $J T_{\sigma}$ deformation. The last option is a TsT in $(X^1,X^+)$, which is similarly easily seen to correspond to the $J T_{\tau}$ deformation as given in~\cite{Frolov:2019xzi}. Of course it is possible to take (linear) combinations of the $J T_{\mu}$ as well as $T\overline{T}$ deformations. Note that $J$ in general can be any (non-necessarily chiral) conserved $\mathfrak{u}(1)$ current.

Various further deformations can be realized via canonical transformations in the light-cone gauge fixing picture of \cite{Frolov:2019xzi}, and many of them can be obviously cast as TsT transformations. However, these would not all be TsT transformations in our natural T dual frame for the $T \overline{T}$ deformation. For example, the $\widetilde{J}T_\mu$, $\mu=\tau$, deformation of ref.~\cite{Frolov:2019xzi} can be naturally viewed as a TsT transformation in $(\widetilde{X}^1,X^+)$, \textit{i.e.}\ it can be viewed as a TsT transformation in a geometry where we have first T-dualized in $X^1$.

\bibliographystyle{JHEP}
\bibliography{refs}

\end{document}